\documentclass[aps,prd,amssymb,nofootinbib,twocolumn,epsf,showpacs]{revtex4}
\usepackage[usenames]{color} 
\usepackage{ulem} 
\usepackage{graphicx}
\usepackage{amssymb}
\usepackage{amsmath}
\usepackage{epstopdf}

\begin{document}

%%%%%%%%%%%%%%%%%%%%%%%%%%%%%%%%%%%%%%%%%%%%%%%%%%%%%%%%%%%%%%%%%%%%%%%%%%%%%%%
% Spectral Approach to the Relativistic Inverse Stellar Structure Problem
%%%%%%%%%%%%%%%%%%%%%%%%%%%%%%%%%%%%%%%%%%%%%%%%%%%%%%%%%%%%%%%%%%%%%%%%%%%%%%%

\title{Spectral Approach to the Relativistic Inverse Stellar 
Structure Problem II}

\author{Lee Lindblom and Nathaniel M. Indik}

\affiliation{Theoretical Astrophysics 350-17, California Institute of
Technology, Pasadena, CA 91125}

\date{\today}

\begin{abstract}
The inverse stellar structure problem determines the equation of state
of the matter in stars from a knowledge of their macroscopic
observables (e.g. their masses and radii).  This problem was solved in
a previous paper by constructing a spectral representation of the
equation of state whose stellar models match a prescribed set of
macroscopic observables.  This paper improves and extends that work in
two significant ways: i) The method is made more robust by accounting
for an unexpected feature of the enthalpy based representations of the
equations of state used in this work.  After making the appropriate
modifications, accurate initial guesses for the spectral parameters
are no longer needed so Monte-Carlo techniques can now be used to ensure
the best fit to the observables.  ii) The method is extended here to
use masses and tidal deformabilities (which will be measured by
gravitational wave observations of neutron-star mergers) as the
macroscopic observables instead of masses and radii.  The accuracy and
reliability of this extended and more robust spectral method is
evaluated in this paper using mock data for observables from stars
based on 34 different theoretical models of the high density
neutron-star equation of state.  In qualitative agreement with
earlier work, these tests suggest the high density part of the
neutron-star equation of state could be determined at the few-percent
accuracy level using high quality measurements of the masses and radii
(or masses and tidal deformabilities) of just two or three neutron
stars.
\end{abstract}


\maketitle

%%%%%%%%%%%%%%%%%%%%%%%%%%%%%%%%%%%%%%%%%%%%%%%%%%%%%%%%%%%%%%%%%%%%%%%%%%%%%%%
\section{Introduction}
\label{s:Introduction}

The purpose of this paper is to improve and extend the spectral
approach to solving the relativistic inverse stellar structure problem
developed in our earlier paper, Lindblom and Indik~\cite{Lindblom12}.
In that approach the density $\epsilon$ and pressure $p$ of the matter
in a particular class of stars (e.g. neutron stars) are represented as
faithful parametric expressions of the form: $\epsilon(h,\gamma_k)$
and $p(h,\gamma_k)$, where $h$ is the enthalpy of the material, and
$\gamma_k$ are parameters that specify the particular equation of
state.  Faithful in this context means that any physical equation of
state has such a representation while every choice of $\gamma_k$
represents a physically possible equation of state (cf.
Lindblom~\cite{Lindblom10}).  Given a specific equation of state in
this form, it is straightforward to solve the relativistic stellar
structure equations to construct stellar models and their macroscopic
observables, e.g. their masses $M(h_c,\gamma_k)$ and radii
$R(h_c,\gamma_k)$.  These macroscopic observables depend on the
equation of state through the parameters $\gamma_k$, as well as the
central enthalpy $h_c$ (or equivalently the central pressure or
density) of the particular stellar model.  Our approach to the inverse
stellar structure problem~\cite{Lindblom12} determines the equation of
state by adjusting the parameters $\gamma_k$ (and $h_c^i$) in the
model observables, e.g. $M(h_c^i,\gamma_k)$ and $R(h_c^i,\gamma_k)$,
to match a set of prescribed values of those observables, e.g. $M_i$
and $R_i$.

The spectral approach to the relativistic inverse stellar structure
problem (summarized above) was tested in our first paper, Lindblom and
Indik~\cite{Lindblom12}, using mock observational data, $M_i$ and
$R_i$, constructed from 34 different theoretical models of the highest
density part of the neutron-star equation of state.  Sequences of
approximate solutions to this problem were constructed by determining
the spectral parameters $\gamma_k$ that minimize the quantity $\chi^2$
defined by,
\begin{eqnarray}
\chi^2(\gamma_k,h_c^i)&=&
\frac{1}{N_\mathrm{stars}}\sum_{i=1}^{N_\mathrm{stars}}\left\{
\left[\log\left(\frac{M(h_c^i,\gamma_k)}{M_i}\right)\right]^2\right.\nonumber\\
&&\qquad\qquad\quad
+\left.\left[\log\left(\frac{R(h_c^i,\gamma_k)}{R_i}\right)\right]^2\right\}.
\qquad
\label{e:ChiSquareDef}
\end{eqnarray}
The accuracies of the resulting spectral equations of state were then
evaluated by comparing with the exact equations of state.  Those tests
showed that the spectral equations of state provide good approximate
solutions to the relativistic inverse stellar structure problem, with
(average) error levels of just a few percent using (mock)
observational data from only two or three stars.  These tests also
showed that the accuracy of the approximations got better (on average)
when more data were used and more spectral parameters were fixed by
the data.

Unfortunately, our implementation of the spectral approach described
above had a serious flaw.  The method worked very well if the search
for the minimum of $\chi^2(\gamma_k,h_c^i)$ in
Eq.~(\ref{e:ChiSquareDef}) began with a reasonably accurate initial
estimate for the spectral parameters $\gamma_k$.  Without an accurate
initial guess, however, the code used to solve this non-linear least
squares problem often crashed.  This flaw made it impossible to
perform searches for the true global minimum of
$\chi^2(\gamma_k,h_c^i)$, or to investigate the structure of that
minimum (in $\gamma_k$ parameter space).  One of the main objectives
of this paper is to understand the cause of this problem, and to use
this understanding to develop a more robust implementation of the
spectral approach.  The root problem turned out to be a subtle and
unexpected feature of the enthalpy based representations of the
equations of state.  This feature is described in some detail in
Sec.~\ref{s:ImprovingMethod}, along with the changes in our initial
implementation of the spectral approach to the inverse stellar
structure problem needed to accommodate it.  Using the resulting more
robust approach, Sec.~\ref{s:ImprovingMethod} contains a more thorough
and systematic study of the mathematical convergence of the sequence
of approximate spectral equations of state produced by this method.

Our analysis of the relativistic inverse stellar structure problem up
to this point has assumed that the masses and radii of neutron stars
would be the first observables measured accurately.  This may turn out
to be the case, but the spectral approach for solving this problem
does not (in principle) depend very strongly on exactly which
observables are used.  Recent work~\cite{Hinderer2008a, Hinderer2008,
  Hinderer2009, Read:2009yp, Hinderer2010, Lackey2012, Damour2012,
  Bernuzzi2012, Read2013, Lackey2013, Pozzo2013, Maselli2013} has
shown that gravitational-wave observations of binary neutron-star
mergers should provide accurate measurements of the masses and tidal
deformabilities of neutron stars once the advanced LIGO-VIRGO network
of detectors becomes operational (within the next few years).  The
possibility of using this type of observational data to solve the
inverse stellar structure problem is explored in
Sec.~\ref{s:TidalDeformability} of this paper.  A new and more
efficient method for evaluating the tidal deformabilities
$\Lambda(h_c,\gamma_k)$ of neutron-star models is presented in
Appendix~\ref{s:AppendixC}, along with an efficient method for
evaluating its derivatives with respect to the parameters $h_c$ and
$\gamma_k$.  The inverse stellar structure problem is tested in
Sec.~\ref{s:TidalDeformability} with masses and tidal deformabilities
computed from the same catalog of 34 theoretical neutron-star
equations of state used in our previous studies.  These tests show
that the high density part of the neutron-star equation of state could
be determined using precision measurements of the masses and tidal
deformabilities of just two or three neutron stars at about the same
level of accuracy that could be achieved using mass and radius data.

Our analysis of the relativistic inverse stellar structure problem
(begun in our first paper~\cite{Lindblom12} and continued here)
focuses on understanding some of the fundamental mathematical aspects
of this problem.  Is it possible to determine the neutron-star
equation of state exactly from a complete exact knowledge of the
macroscopic observable properties of these stars, i.e., does this
problem have a unique solution?  Can numerical methods can be devised
whose approximate solutions converge to the exact equation of state
when a complete exact knowledge of the macroscopic observables of
these stars is available?  What level of numerical approximation and
how many macroscopic observable data points are needed to achieve
reasonable levels of accuracy for ``realistic'' neutron-star equations
of state?  A number of researchers have studied various observational
and data-analysis questions associated with the inverse stellar
structure problem, both in the context of using mass and radius
observations~\cite{Ozel2009, Steiner2010, Guver2012a, Guver2013,
  Steiner2012}, and in the context of using mass and tidal
deformability measurements from gravitational-wave
observations~\cite{Hinderer2008a, Hinderer2008, Hinderer2009,
  Read:2009yp, Hinderer2010, Lackey2012, Damour2012, Bernuzzi2012,
  Read2013, Lackey2013, Pozzo2013, Maselli2013}.  To our knowledge,
our studies of the more fundamental questions about solving the
inverse stellar structure problem described in our papers are unique.
We discuss in more detail some of the basic differences between our
results and those reported by others in Sec.~\ref{s:Discussion}.

\section{Improving the Method}
\label{s:ImprovingMethod}

The spectral approach to the relativistic inverse stellar structure
problem outlined above requires the use of a faithful parametric
representation of the equation of state.  There are a variety of ways
to construct such representations (cf. Lindblom~\cite{Lindblom10}),
but the most useful for solving the relativistic stellar structure
problem (and its inverse) represent the energy density $\epsilon$ and
pressure $p$ of the stellar matter as functions of the relativistic
enthalpy: $\epsilon(h,\gamma_k)$ and $p(h,\gamma_k)$. The parameters
$\gamma_k$ specify the particular equation of state, and the
relativistic specific enthalpy $h$ is defined by the integral,
\begin{eqnarray}
h(p) = \int_0^p \frac{dp'}{\epsilon(p') + p'}.
\label{e:EnthalpyDef}
\end{eqnarray} 
Representing the equation of state in this way makes it possible to
transform the stellar structure equations into a form that can be
solved numerically more accurately and efficiently than the standard
Oppenheimer-Volkoff form of the equations~\cite{Lindblom1992,
  Lindblom12}.

An important feature of the enthalpy (from the perspective of the
inverse stellar structure problem) is the unexpected diversity of its
high pressure limit: $h_\infty\equiv \lim_{\,p\rightarrow\infty}h(p)$.
This limit is infinite in some equations of state, while it is finite
in others.  For example, an equation of state of the form
$\epsilon=\epsilon_0 + p$, has an enthalpy given by
$h(p)=\log\sqrt{1+2p/\epsilon_0}$ with $h_\infty=\infty$.  However the
equation of state $\epsilon = \epsilon_0\,e^{\,p/p_0} - p$, has an
enthalpy given by $h(p)=p_0(1-e^{-p/p_0})/\epsilon_0$, with $h_\infty=
p_0/\epsilon_0$.  This diversity in $h_\infty$ complicates the problem
of writing a robust code to find the minimum of
$\chi^2(\gamma_k,h_c^i)$.

For any given equation of state, the parameters $h_c^i$ that specify
the central enthalpy of each stellar model must satisfy $h_c^i\leq
h_\infty$.  Since $h_\infty$ depends on the equation of state, these
conditions on $h_c^i$ also depend on the parameters $\gamma_k$ used to
specify the particular equation of state: $h_c^i\leq
h_\infty(\gamma_k)$.  Any algorithm that explores the structure of the
function $\chi^2(\gamma_k,h_c^i)$ to find its minimum, must therefore
ensure that the inequalities $h_c^i\leq h_\infty(\gamma_k)$ are
satisfied at every step of the process.

We assumed (implicitly) in our original implementation of the spectral
approach that $h_\infty=\infty$, so it seemed unnecessary to check the
conditions $h_c^i\leq h_\infty(\gamma_k)$.  This error is benign
whenever the initial choices for the parameters $\gamma_k$ and $h_c^i$
are close to a minimum where the conditions are satisfied.  However,
this limitation prevented us from exploring the structure of
$\chi^2(\gamma_k,h_c^i)$ except in the immediate neighborhood of a
good initial estimate.  Whenever the condition $h_c^i\leq
h_\infty(\gamma_k)$ was violated for some reason, our original code
produced unpredictable results: sometimes generating unphysical
(e.g. negative) densities, and sometimes simply crashing.  This
limitation therefore prevented us from using Monte Carlo methods to
explore the $\gamma_k$ and $h_c^i$ parameter space more widely, and
made it impossible to determine whether any particular local minimum
of $\chi^2(\gamma_k,h_c^i,)$ was also its global minimum.

The minima of complicated non-linear functions like
$\chi^2(\gamma_k,h_c^i)$ are generally found numerically using
iterative methods.  At an abstract level these methods begin with some
choice of the parameters which are then refined in some way to produce
an estimate that is closer to a minimum.  This process is repeated
until an appropriate convergence criterion is satisfied.  At each step
in this process the parameters must satisfy $h_c^i\leq
h_\infty(\gamma_k)$, or the code which evaluates
$\epsilon(h,\gamma_k,)$ and $p(h,\gamma_k)$ will fail whenever $h$
enters the range $h_\infty\leq h\leq h_c^i$.  The upper limit on the
range of physical enthalpies $h_\infty(\gamma_k)$ must therefore be
re-evaluated at each step that changes the values of the spectral
parameters $\gamma_k$.  Appendix~\ref{s:AppendixA} describes in detail
how the value of a good estimate $h_{\max}\leq h_\infty(\gamma_k)$ can
be determined for the spectral equations of state used in our
approach.  The conditions $h_c^i\leq h_{\max}$ are then checked at
each step of the iterative process that finds the minimum of
$\chi^2(\gamma_k,h_c^i)$.  If any of the $h_c^i$ violate this
condition at any step, then all the $h_c^i$ at this step are scaled
(down) so the conditions $h_c^i\leq h_{\max}$ are satisfied before
proceeding.  Testing and re-scaling the $h_c^i$ (if necessary) at each
step is the biggest improvement in our new more robust implementation
of the spectral approach to the inverse stellar structure problem.
With this change it becomes possible to use Monte Carlo methods to
explore the global minimum of $\chi^2(\gamma_k,h_c^i)$.

This new improved implementation of the spectral approach to the
relativistic inverse stellar structure problem has been tested using
mock observational data for the masses and radii based on the 34
theoretical high-density neutron-star equations of state.  These mock
data sets consist of $N_\mathrm{stars}$ $[M_i,R_i]$ data pairs, with
the masses uniformly spaced between $1.2M_\odot$ (a typical minimum
mass for astrophysical neutron-stars) and the maximum mass
$M_\mathrm{max}$ for each theoretical equation of state.  See Read,
et al.~\cite{Read:2008iy} for descriptions of these 34 theoretical
equations of state used in our tests, along with citations to the
original nuclear physics papers on which they are based.  

The mock data used here differ in only two minor ways from those used
in our original work~\cite{Lindblom12}.  First, the method of
extrapolating above and below the highest and lowest entries in those
tabulated theoretical equations of state was changed slightly for
these new tests.  The new versions of our interpolation and
extrapolation formulas are given in detail in
Appendix~\ref{s:AppendixB}, while the old version is described in
Appendix B of Ref.~\cite{Lindblom12}.  The second change made some
(minor) corrections to some of the theoretical equation of state
tables.  In particular we found that some of the tabulated equations
of state were non-monotonic (and therefore non-physical) at a density
of about $1.67\times 10^{12}\, g/cm^3$.  The effected equations of
state were: APR1, APR2, APR3, APR4, ENG, H1, H2, H3, MPA1, MS1B, MS1,
PCL2, PS, WFF1, WFF2, and WFF3.  We corrected these problems simply by
removing the one row in each table at the density where this
non-monotonicity occurred.  The resulting interpolated equations of
state are then monotonic. The result of these two minor changes made
it possible to compute stellar models and their observational
properties based on these tabulated equations of state more accurately
and reliably.

In these tests of our new improved implementation of the spectral
approach to the inverse stellar structure problem, we begin the
calculation of the minimum of $\chi(\gamma_k,h_c^i)$ by choosing a
good initial estimate for the parameters $\gamma_k$ and $h_c^i$.  We
refine this initial estimate using the Levenberg-Marquardt
algorithm~\cite{numrec_f} to find a local minimum of
$\chi(\gamma_k,h_c^i)$.  Once completed, we explore the neighborhood
of this minimum by adding small random changes to each of the
parameters $\gamma_k$ and $h_c^i$.  The minimum of
$\chi(\gamma_k,h_c^i)$ is then recomputed using Levenberg-Marquardt
with these randomized initial parameter values.  This process is
repeated until a minimum is found with
$\chi(\gamma_k,h_c^i)<10^{-10}$, or until 100 subsequent randomized
steps fail to reduce the smallest minimum further.

The results of these tests are summarized in Table~\ref{t:TableI}.
For each equation of state the inverse stellar structure problem has
been solved by fitting $N_{\gamma_k}$ different spectral parameters to
mock data sets containing $N=N_\mathrm{stars}=N_{\gamma_k}$ pairs of
mass $M_i$ and radius $R_i$ data.  The minimum value of the fitting
function $\chi_{N}$ is given for each of these solutions in
Table~\ref{t:TableI}.  Two additional quantities, $\Delta^{MR}_{N}$
and $\Upsilon^{MR}_{N}$ are also included in Table~\ref{t:TableI} that
measure how accurately the $N$ parameter spectral equation of state
agrees with the original used to compute the mock mass-radius
observables.  The function $\Delta^{MR}_{N}$ is defined by:
\begin{eqnarray}
\left(\Delta^{MR}_{N}\right)^2=\frac{1}{N_\mathrm{eos}}
\sum_{i=1}^{N_\mathrm{eos}}\left[\log\left(\frac{\epsilon(h_i,\gamma_k)}{\epsilon_i}
\right)\right]^2.
\label{e:DeltaMR_Def}
\end{eqnarray}
The sum in Eq.~(\ref{e:DeltaMR_Def}) is over the points,
$[\epsilon_i,h_i]$ from the tabulated theoretical equation of state
table.  Only the $N_\mathrm{eos}$ points that lie in the range
$h_0\leq h_i \leq \max h_c$ are included in this sum, where $h_0$ is
the lower limit of the spectral domain, and $\max h_c$ is the central
value of $h$ in the maximum mass neutron star for this equation of
state.  The quantity $\Delta^{MR}_{N}$ therefore measures the average
error in the spectral part of the equation of state [i.e., the part
  with densities above $\epsilon(h_0)$] that occur within neutron
stars.\footnote{We follow the convention used in Read, et
  al.~\cite{Read:2008iy} and in Lindblom and Indik~\cite{Lindblom12}
  and choose the density $\epsilon(h_0)$ at the lower end of the
  spectral domain to be about half nuclear density.}  The best
possible spectral fit to each of these theoretical neutron-star
equations of state was determined in Ref.~\cite{Lindblom10}, and the
average errors $\Delta^{EOS}_{N}$ of those best $N_{\gamma_k}$
parameter spectral fits are given in Table II of that reference.  The
quantity $\Upsilon^{MR}_{N}$ measures the relative accuracy between
the $N$ parameter spectral equation of state determined by solving the
inverse stellar structure problem, and the best possible spectral fit:
\begin{eqnarray}
\Upsilon^{MR}_{N}=\frac{\Delta^{MR}_{N}}{\Delta^{EOS}_{N}}.
\label{e:UpsilonMR_Def}
\end{eqnarray}

Except for the improvements described above, the tests performed here
are identical to those performed in our original implementation of the
spectral approach.  The new results given in Table~\ref{t:TableI} are
therefore directly comparable to those given in Table I of Lindblom
and Indik~\cite{Lindblom12}.  The most obvious differences between the
two tables are the values of $\chi_N$.  All of the new $\chi_N$
(except one) are less than our convergence criterion, $\chi_N<
10^{-10}$, while in contrast very few of the original $\chi_N$ were
able to meet this condition.  These improvements in the values of
$\chi_N$ are due (mostly) to the use of Monte Carlo methods to ensure
that a global rather than just a local minimum of
$\chi^2(\gamma_k,h_c^i)$ is obtained.
\begin{table*}[!htb]
\begin{center}
\caption{Accuracies of the neutron-star equations of state obtained by
  solving the inverse stellar structure problem using mass-radius
  data.  $\Delta_N^{MR}$ measures the average fractional error of the
  equation of state obtained by fitting to $N$ different
  $[M_i,R_i]$ data pairs.  The parameter $\Upsilon_N^{MR}$ measures
  the ratio of $\Delta_N^{MR}$ to the errors in the optimal
  $N$-parameter spectral fit to each equation of state.  The parameter
  $\chi_N$ measures the accuracy with which the model masses
  $M(h_c^i,\gamma_k)$ and radii $R(h_c^i,\gamma_k)$ produced by the
  approximate spectral equation of state match the exact $M_i$
  and $R_i$ data.
\label{t:TableI}}
\medskip
\begin{tabular}{|l|cccc|cccc|cccc|}
\hline\hline
    EOS   &$\Delta_{2}^{MR}$ &$\Delta_{3}^{MR}$ &$\Delta_{4}^{MR}$ 
     &$\Delta_{5}^{MR}$ &$\Upsilon_2^{MR}$ &$\Upsilon_3^{MR}$ 
     &$\Upsilon_4^{MR}$ &$\Upsilon_5^{MR}$ 
     &$\chi_2$ &$\chi_3$ &$\chi_4$ &$\chi_5$ \\
\hline
    PAL6 &  0.0034 &  0.0018 &  0.0007 &  0.0003 &  1.06 &  1.09 &  1.33 &  1.91 & $ 9.2\times 10^{-12}$ & $ 4.1\times 10^{-11}$ & $ 5.1\times 10^{-11}$ & $ 5.2\times 10^{-11}$ \\
     SLy &  0.0107 &  0.0040 &  0.0022 &  0.0011 &  1.17 &  1.13 &  1.30 &  1.68 & $ 4.2\times 10^{-11}$ & $ 5.3\times 10^{-11}$ & $ 7.9\times 10^{-11}$ & $ 8.3\times 10^{-11}$ \\
    APR1 &  0.0746 &  0.0422 &  0.0314 &  0.0172 &  1.05 &  1.27 &  1.68 &  2.10 & $ 4.1\times 10^{-11}$ & $ 2.2\times 10^{-11}$ & $ 8.8\times 10^{-11}$ & $ 9.4\times 10^{-11}$ \\
    APR2 &  0.0313 &  0.0165 &  0.0094 &  0.0068 &  1.01 &  1.18 &  1.49 &  2.02 & $ 3.9\times 10^{-11}$ & $ 8.8\times 10^{-11}$ & $ 3.2\times 10^{-11}$ & $ 7.2\times 10^{-11}$ \\
    APR3 &  0.0266 &  0.0061 &  0.0030 &  0.0022 &  1.06 &  1.13 &  1.24 &  1.49 & $ 3.2\times 10^{-11}$ & $ 2.4\times 10^{-11}$ & $ 9.4\times 10^{-11}$ & $ 9.8\times 10^{-11}$ \\
    APR4 &  0.0258 &  0.0037 &  0.0017 &  0.0016 &  1.03 &  1.23 &  1.26 &  1.16 & $ 5.6\times 10^{-11}$ & $ 3.4\times 10^{-11}$ & $ 7.7\times 10^{-11}$ & $ 8.1\times 10^{-11}$ \\
     FPS &  0.0047 &  0.0061 &  0.0096 &  0.0049 &  1.06 &  1.44 &  2.53 &  2.69 & $ 2.6\times 10^{-11}$ & $ 3.7\times 10^{-11}$ & $ 7.5\times 10^{-11}$ & $ 8.3\times 10^{-11}$ \\
    WFF1 &  0.0552 &  0.0169 &  0.0220 &  0.0158 &  1.04 &  1.59 &  3.19 &  2.41 & $ 9.6\times 10^{-11}$ & $ 6.0\times 10^{-11}$ & $ 6.6\times 10^{-11}$ & $ 9.6\times 10^{-11}$ \\
    WFF2 &  0.0277 &  0.0146 &  0.0084 &  0.0055 &  1.01 &  1.21 &  1.18 &  1.46 & $ 3.4\times 10^{-11}$ & $ 6.4\times 10^{-11}$ & $ 7.8\times 10^{-11}$ & $ 9.5\times 10^{-11}$ \\
    WFF3 &  0.0127 &  0.0147 &  0.0124 &  0.0110 &  1.14 &  1.43 &  2.09 &  1.98 & $ 3.0\times 10^{-11}$ & $ 4.0\times 10^{-11}$ & $ 7.1\times 10^{-11}$ & $ 8.8\times 10^{-11}$ \\
    BBB2 &  0.0332 &  0.0328 &  0.0303 &  0.0131 &  1.01 &  1.14 &  1.39 &  1.42 & $ 1.2\times 10^{-11}$ & $ 4.4\times 10^{-11}$ & $ 9.6\times 10^{-11}$ & $ 6.8\times 10^{-11}$ \\
  BPAL12 &  0.0181 &  0.0107 &  0.0068 &  0.0075 &  1.06 &  1.08 &  1.37 &  3.36 & $ 4.6\times 10^{-12}$ & $ 1.3\times 10^{-11}$ & $ 4.6\times 10^{-11}$ & $ 5.6\times 10^{-11}$ \\
     ENG &  0.0204 &  0.0247 &  0.0201 &  0.0478 &  1.01 &  1.33 &  1.36 &  4.25 & $ 3.6\times 10^{-11}$ & $ 5.6\times 10^{-11}$ & $ 7.9\times 10^{-11}$ & $ 9.7\times 10^{-11}$ \\
    MPA1 &  0.0328 &  0.0040 &  0.0049 &  0.0057 &  1.27 &  1.23 &  1.60 &  2.50 & $ 7.8\times 10^{-11}$ & $ 3.6\times 10^{-11}$ & $ 2.3\times 10^{-11}$ & $ 7.4\times 10^{-11}$ \\
     MS1 &  0.0474 &  0.0157 &  0.0132 &  0.0009 &  1.65 &  2.77 &  3.63 &  2.49 & $ 5.6\times 10^{-11}$ & $ 5.6\times 10^{-11}$ & $ 6.6\times 10^{-11}$ & $ 8.9\times 10^{-11}$ \\
     MS2 &  0.0159 &  0.0044 &  0.0009 &  0.0006 &  1.35 &  1.86 &  2.17 &  3.41 & $ 1.3\times 10^{-15}$ & $ 8.6\times 10^{-16}$ & $ 1.3\times 10^{-15}$ & $ 1.1\times 10^{-15}$ \\
    MS1B &  0.0305 &  0.0149 &  0.0084 &  0.0017 &  1.53 &  2.32 &  2.85 &  6.08 & $ 6.6\times 10^{-11}$ & $ 6.3\times 10^{-11}$ & $ 9.4\times 10^{-11}$ & $ 9.3\times 10^{-11}$ \\
      PS &  0.1047 &  0.0779 &  0.1125 &  0.0432 &  1.67 &  2.59 &  3.74 &  2.58 & $ 6.9\times 10^{-11}$ & $ 7.8\times 10^{-11}$ & $ 5.7\times 10^{-11}$ & $ 1.5\times 10^{-5}$ \\
     GS1 &  0.0965 &  0.0604 &  0.0388 &  0.0445 &  1.08 &  1.56 &  1.03 &  1.78 & $ 5.1\times 10^{-12}$ & $ 2.0\times 10^{-12}$ & $ 6.1\times 10^{-12}$ & $ 1.8\times 10^{-11}$ \\
     GS2 &  0.0885 &  0.0888 &  0.1144 &  0.0426 &  1.46 &  2.02 &  2.63 &  1.35 & $ 3.2\times 10^{-11}$ & $ 6.9\times 10^{-11}$ & $ 6.7\times 10^{-11}$ & $ 8.0\times 10^{-11}$ \\
  BGN1H1 &  0.1352 &  0.1702 &  0.1356 &  0.1382 &  1.54 &  3.40 &  3.06 &  3.94 & $ 3.4\times 10^{-11}$ & $ 5.2\times 10^{-11}$ & $ 6.4\times 10^{-11}$ & $ 9.2\times 10^{-11}$ \\
    GNH3 &  0.0174 &  0.0183 &  0.0389 &  0.0171 &  1.27 &  1.92 &  4.72 &  2.93 & $ 8.5\times 10^{-12}$ & $ 3.0\times 10^{-11}$ & $ 5.7\times 10^{-11}$ & $ 9.5\times 10^{-11}$ \\
      H1 &  0.0294 &  0.0161 &  0.0127 &  0.0105 &  1.44 &  1.29 &  1.47 &  1.45 & $ 4.5\times 10^{-11}$ & $ 4.1\times 10^{-11}$ & $ 7.2\times 10^{-11}$ & $ 9.3\times 10^{-11}$ \\
      H2 &  0.0211 &  0.0279 &  0.0146 &  0.0221 &  1.19 &  2.01 &  2.12 &  3.22 & $ 1.7\times 10^{-11}$ & $ 6.4\times 10^{-11}$ & $ 4.3\times 10^{-11}$ & $ 8.5\times 10^{-11}$ \\
      H3 &  0.0139 &  0.0201 &  0.0176 &  0.0097 &  1.09 &  1.79 &  2.08 &  1.39 & $ 3.1\times 10^{-11}$ & $ 3.8\times 10^{-11}$ & $ 9.7\times 10^{-11}$ & $ 9.9\times 10^{-11}$ \\
      H4 &  0.0132 &  0.0259 &  0.0187 &  0.0105 &  1.28 &  2.60 &  2.76 &  1.56 & $ 8.3\times 10^{-11}$ & $ 6.1\times 10^{-11}$ & $ 5.7\times 10^{-11}$ & $ 8.5\times 10^{-11}$ \\
      H5 &  0.0140 &  0.0296 &  0.0118 &  0.0160 &  1.02 &  2.21 &  2.00 &  3.25 & $ 3.6\times 10^{-11}$ & $ 1.0\times 10^{-10}$ & $ 6.4\times 10^{-11}$ & $ 8.7\times 10^{-11}$ \\
      H6 &  0.0150 &  0.0141 &  0.0205 &  0.0157 &  1.09 &  1.03 &  1.57 &  1.38 & $ 9.1\times 10^{-11}$ & $ 9.8\times 10^{-11}$ & $ 7.9\times 10^{-11}$ & $ 1.0\times 10^{-10}$ \\
      H7 &  0.0134 &  0.0212 &  0.0124 &  0.0129 &  1.09 &  1.88 &  2.17 &  2.28 & $ 1.9\times 10^{-11}$ & $ 8.3\times 10^{-11}$ & $ 9.4\times 10^{-11}$ & $ 8.5\times 10^{-11}$ \\
    PCL2 &  0.0374 &  0.0152 &  0.0101 &  0.0250 &  1.35 &  1.16 &  1.04 &  3.06 & $ 4.8\times 10^{-11}$ & $ 6.4\times 10^{-11}$ & $ 2.6\times 10^{-11}$ & $ 9.3\times 10^{-11}$ \\
    ALF1 &  0.0796 &  0.0664 &  0.1040 &  0.0768 &  1.08 &  1.39 &  2.59 &  2.70 & $ 6.8\times 10^{-11}$ & $ 5.6\times 10^{-11}$ & $ 9.3\times 10^{-11}$ & $ 9.0\times 10^{-11}$ \\
    ALF2 &  0.0723 &  0.0598 &  0.0485 &  0.0218 &  1.04 &  1.21 &  1.75 &  1.22 & $ 5.8\times 10^{-11}$ & $ 8.1\times 10^{-11}$ & $ 8.6\times 10^{-11}$ & $ 9.3\times 10^{-11}$ \\
    ALF3 &  0.0404 &  0.0178 &  0.0202 &  0.1229 &  1.04 &  1.19 &  1.43 &  9.13 & $ 2.2\times 10^{-11}$ & $ 5.2\times 10^{-11}$ & $ 8.8\times 10^{-11}$ & $ 3.1\times 10^{-11}$ \\
    ALF4 &  0.0839 &  0.0182 &  0.0218 &  0.0394 &  1.18 &  1.35 &  2.19 &  4.15 & $ 7.6\times 10^{-11}$ & $ 5.2\times 10^{-11}$ & $ 9.1\times 10^{-11}$ & $ 9.2\times 10^{-11}$ \\
\hline
Averages &  0.0396 &  0.0289 &  0.0276 &  0.0239 &  1.22 &  1.65 &  2.14 &  2.77 & &&&\\
\hline\hline
 \end{tabular}
\end{center}
\end{table*}

The parameters $\Delta_N^{MR}$ in Table~\ref{t:TableI} that quantify
the errors in the spectral equations of state are slightly larger (on
average) than those obtained using our original implementation of the
method.  The averages of these quantities (over the 34 different
theoretical equations of state) in the new tests are
$\Delta_2^{MR}=0.040$, $\Delta_3^{MR}=0.029$, $\Delta_4^{MR}=0.028$,
$\Delta_5^{MR}=0.024$, while the values found in the original tests
were $\Delta_2^{MR}=0.040$, $\Delta_3^{MR}=0.029$,
$\Delta_4^{MR}=0.023$, $\Delta_5^{MR}=0.017$, The errors in the fits
with $N_{\gamma_k}=2$ and $N_{\gamma_k}=3$ are almost identical to
those from the original tests.  But the errors in the fits with
$N_{\gamma_k}=4$ and $N_{\gamma_k}=5$ are slightly larger.  The basic
reason for these differences comes from the simple fact that the
original method used good initial estimates of the parameters
$\gamma_k$ and $h_c^i$, followed by Levenberg-Marquardt minimization
to find the nearest minimum.  This local minimum was not always the
global minimum of $\chi^2(\gamma_k,h_c^i)$, and in some cases
(especially for larger values of $N_{\gamma_k}$) the real global
minimum has somewhat larger equation of state errors than the local
minimum.  Despite these increases, however, the improved method still
provides very good approximations to the neutron-star equation of
state: i.e.,  average accuracy levels of just a few percent are
achieved using using high precision (mock) observational data from just
two or three stars.

In a few cases, the equation of state errors $\Delta_N^{MR}$ and
$\Upsilon_N^{MR}$ in Table~\ref{t:TableI} are much larger than the
values found using our original methods in Ref.~\cite{Lindblom12}.  In
these cases the error quantities appear non-convergent as the number
of parameters $N_{\gamma_k}$ is increased.  We now believe that the
least squares method itself may be responsible for some of these
failures.  It is well known, for example, that interpolating
polynomials constructed by least squares fits to data at equally
spaced points are unreliable when $N^2>4 K$, where $N$ is the order of
the polynomial fit and $K$ the number of data points, cf \S 4.3.4 of
Dahlquist and Bj\"orck~\cite{Dahlquist2003}.  When $N$ exceeds this
amount, the least squares method tends to produce fits that accurately
pass through the $K$ fixed data points, but oscillate wildly about the
true solution between these points.  This is referred to in the
literature as the Runge phenomenon.  While the particular non-linear
least squares minimization used in our spectral method is not
strictly equivalent to polynomial interpolation, our expectation is
that our method probably exhibits some form of Runge phenomenon unless
appropriate restrictions are made on the number of spectral
parameters, i.e., some condition of the form
$N_{\gamma_k}<F(N_\mathrm{stars})$.
\begin{figure}[tbp!]
%\centerline{\includegraphics[width=3in]{pal6_delta_conv.eps}}
\centerline{\includegraphics[width=3in]{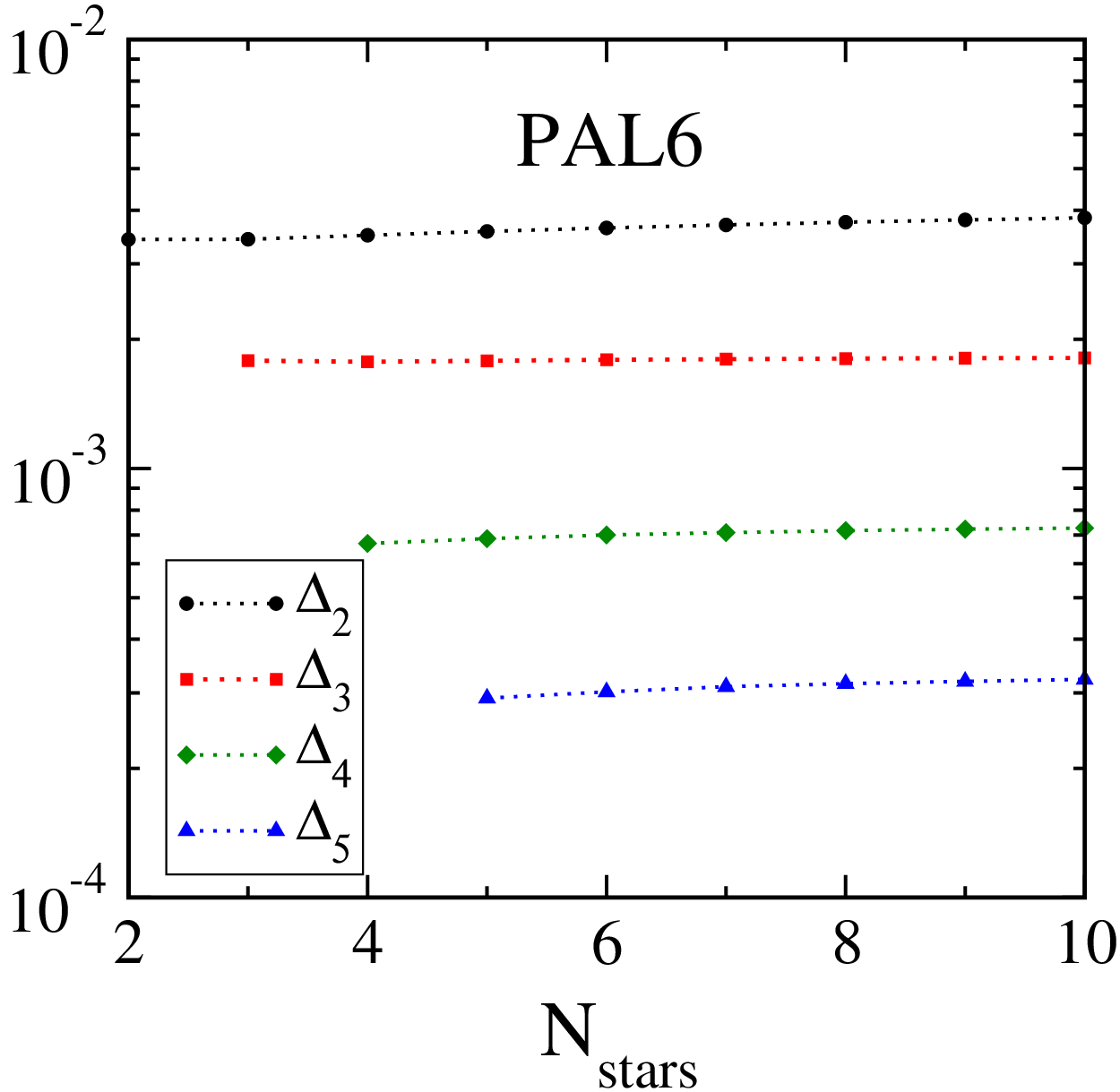}}
\caption{\label{f:pal6_delta_conv} Equation of state errors
  $\Delta_{N_{\gamma_k}}^{MR}$ as functions of the number of
  mass-radii data points, $N_\mathrm{stars}$, used to fix the spectral
  parameters $\gamma_k$ in an $N_{\gamma_k}$ parameter spectral
  expansion.  These results use mass-radius data computed with the PAL6
  equation of state.}
\end{figure}

\begin{figure}[tbp!]
%\centerline{\includegraphics[width=3in]{bgn1h1_delta_conv.eps}}
\centerline{\includegraphics[width=3in]{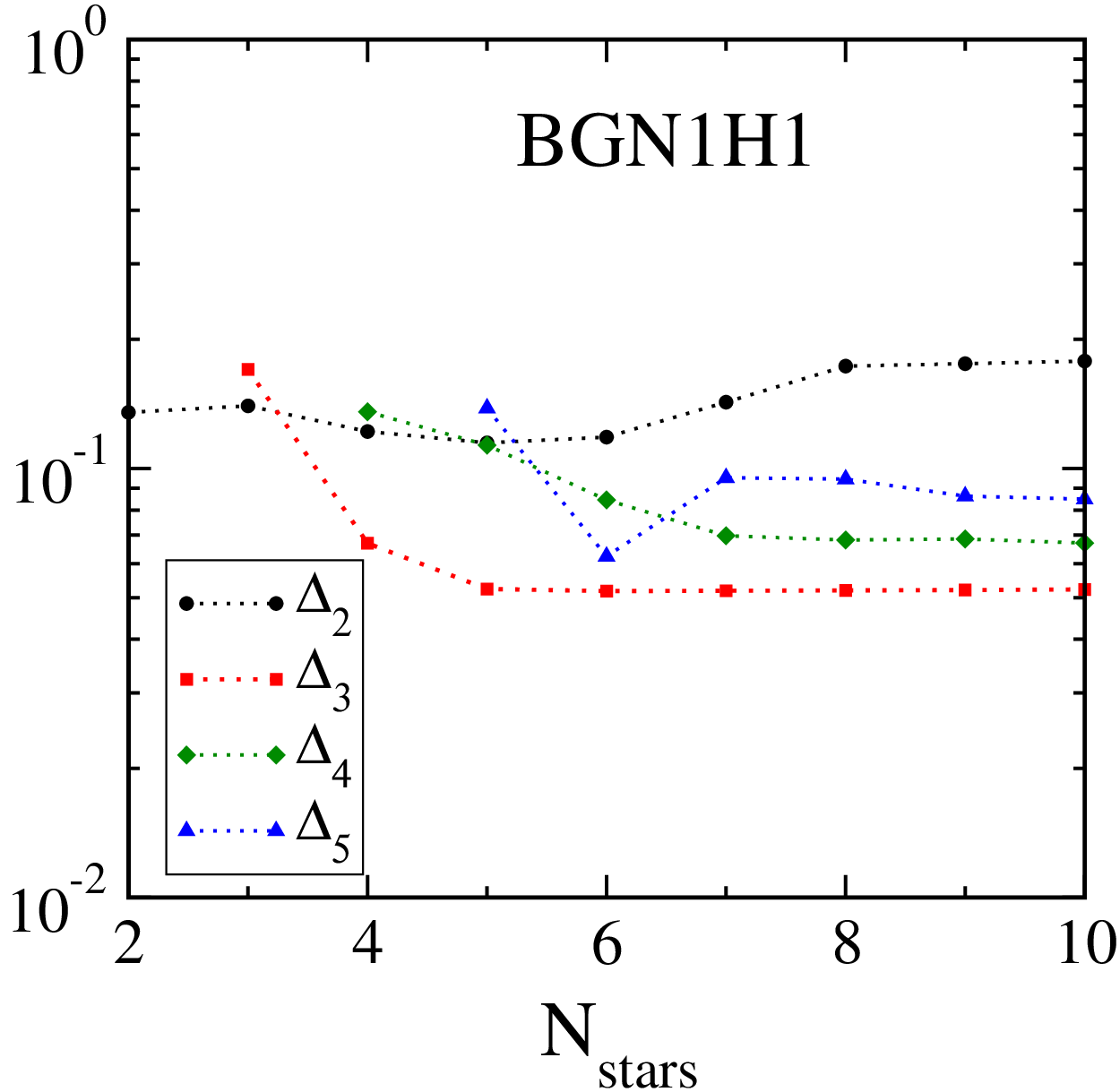}}
\caption{\label{f:bgn1h1_delta_conv} Equation of state errors
  $\Delta_{N_{\gamma_k}}^{MR}$ as functions of the number of
  mass-radii data points, $N_\mathrm{stars}$, used to fix the spectral
  parameters $\gamma_k$ in an $N_{\gamma_k}$ parameter spectral
  expansion.  These results use mass-radius data computed with the
  BGN1H1 equation of state.}
\end{figure}

At present we do not know an analytical expression for the function
$F(N_\mathrm{stars})$ that determines this stability criterion, but
we can explore this question by examining the numerical convergence of
our spectral equations of state.  To do that we have examined in more
detail the spectral solutions using mock observational data
constructed from the PAL6 and the BGN1H1 equations of state.  These
cases represent the best (PAL6) and the worst (BGN1H1) spectral
representations of the 34 equations of state used in our
tests~\cite{Lindblom10, Lindblom12}.  Figures~\ref{f:pal6_delta_conv}
and \ref{f:bgn1h1_delta_conv} show the dependence of the error
quantities $\Delta_{N_{\gamma_k}}^{MR}$ for these cases as functions
of the number of data points $N_\mathrm{stars}$ used in the solution.
The results in the best case, Fig.~\ref{f:pal6_delta_conv}, show the
exponential spectral convergence that is expected in the high $N$
limit.  There are no significant changes in
$\Delta_{N_{\gamma_k}}^{MR}(N_\mathrm{stars})$ as $N_\mathrm{stars}$ is
increased above the minimum $N_\mathrm{stars}=N_{\gamma_k}$, and
$\Delta_{N_{\gamma_k}}^{MR}$ decreases exponentially as $N_{\gamma_k}$
increases.  The worst case, Fig.~\ref{f:bgn1h1_delta_conv}, shows
definite signs of the Runge phenomenon.  The error functions
$\Delta_{N_{\gamma_k}}^{MR}(N_\mathrm{stars})$ for fixed $N_{\gamma_k}$ in
this case decrease significantly as $N_\mathrm{stars}$ increases.  The
BGN1H1 equation of state has a strong phase transition in the density
range where the spectral methods are used, so it is not really
surprising that even in the large $N_\mathrm{stars}$ limit the
spectral equations of state in this case have yet to enter the
convergent range for the relatively small values of $N_{\gamma_k}$
used in these tests.  The good news is that even in this terrible
case, the errors in the inferred spectral equations of state are never
worse than about 20\%, and it appears that results in the 5--10\%
range can be obtained using high quality observational data from about
six stars.

\begin{figure}[tbp!]
%\centerline{\includegraphics[width=3in]{ps_delta_conv.eps}}
\centerline{\includegraphics[width=3in]{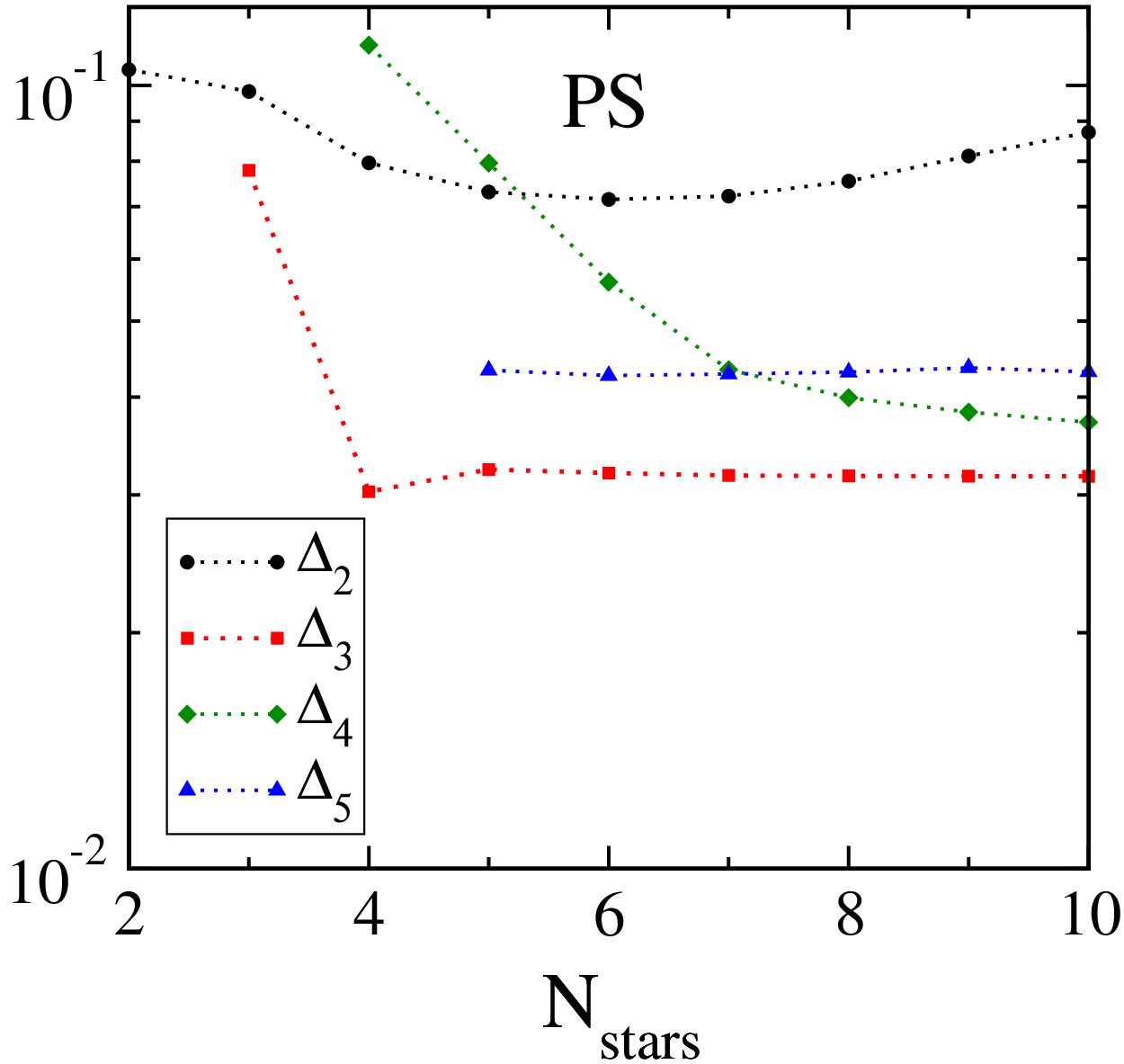}}
\caption{\label{f:ps_delta_conv} Equation of state errors
  $\Delta_{N_{\gamma_k}}^{MR}$ as functions of the number of
  mass-radii data points, $N_\mathrm{stars}$, used to fix the spectral
  parameters $\gamma_k$ in an $N_{\gamma_k}$ parameter spectral
  expansion.  These results use mass-radius data computed with the PS
  equation of state.}
\end{figure}

\begin{figure}[htbp!]
%\centerline{\includegraphics[width=3in]{gs2_delta_conv.eps}}
\centerline{\includegraphics[width=3in]{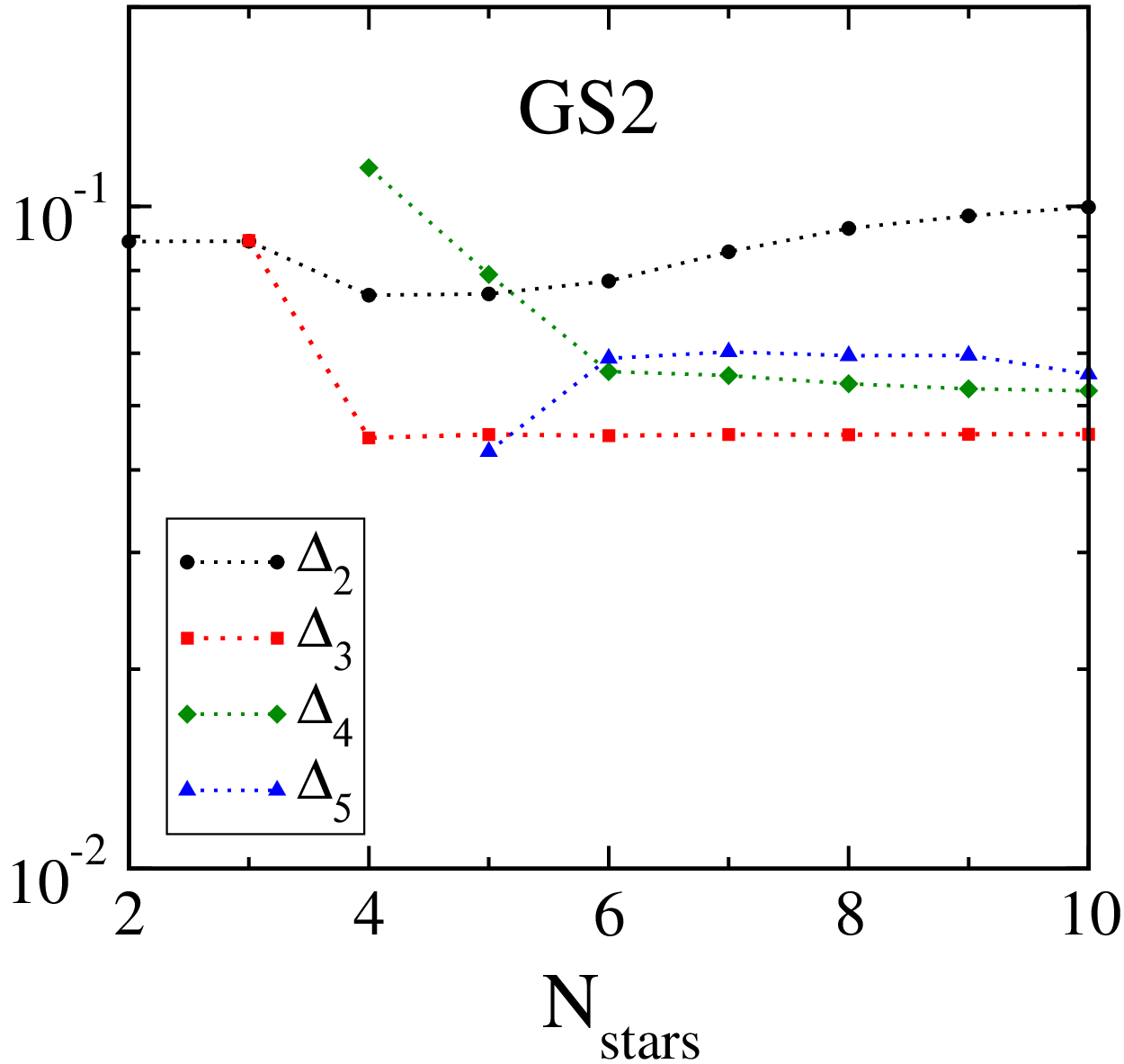}}
\caption{\label{f:gs2_delta_conv} 
Equation of state errors $\Delta_{N_{\gamma_k}}^{MR}$ as functions
of the number of mass-radii data points, $N_\mathrm{stars}$,
used to fix the spectral parameters $\gamma_k$ in
an $N_{\gamma_k}$ parameter spectral expansion.  These results
use mass-radius data computed with the GS2 equation
of state.}
\end{figure}

We have also examined the numerical convergence of our spectral fits
in more detail for several additional cases that show significant
deviations from ideal convergence: PS, GS2, ALF1, and ALF3.  The
sequences of error measures $\Delta_{N_{\gamma_k}}^{MR}$ given in
Table~\ref{t:TableI} clearly appear to be non-convergent for those
cases.  The PS equation of state is also anomalous because it is the
only case where our method fails to find a minimum of
$\chi^2(\gamma_k,h_c^i)$ satisfying our convergence criterion:
$\chi(\gamma_k,h_c^i)\leq 10^{-10}$.  Figures~\ref{f:ps_delta_conv}
and \ref{f:gs2_delta_conv} show the error quantities
$\Delta_{N_{\gamma_k}}^{MR}$ for the PS and the GS2 cases as functions of
the number of data points $N_\mathrm{stars}$ used to construct the
solutions.  These cases both show definite signs of the Runge
phenomenon: the error functions
$\Delta_{N_{\gamma_k}}^{MR}(N_\mathrm{stars})$ for fixed $N_{\gamma_k}$
decrease significantly as $N_\mathrm{stars}$ increases.  So the
unexpectedly large values of $\Delta_{N_{\gamma_k}}^{MR}(N_\mathrm{stars})$
seen in the $N_{\gamma_k}=N_\mathrm{stars}$ solutions reported in
Table~\ref{t:TableI} for those cases are in fact anomalous.

\begin{figure}[htbp!]
%\centerline{\includegraphics[width=3in]{alf3_delta_conv.eps}}
\centerline{\includegraphics[width=3in]{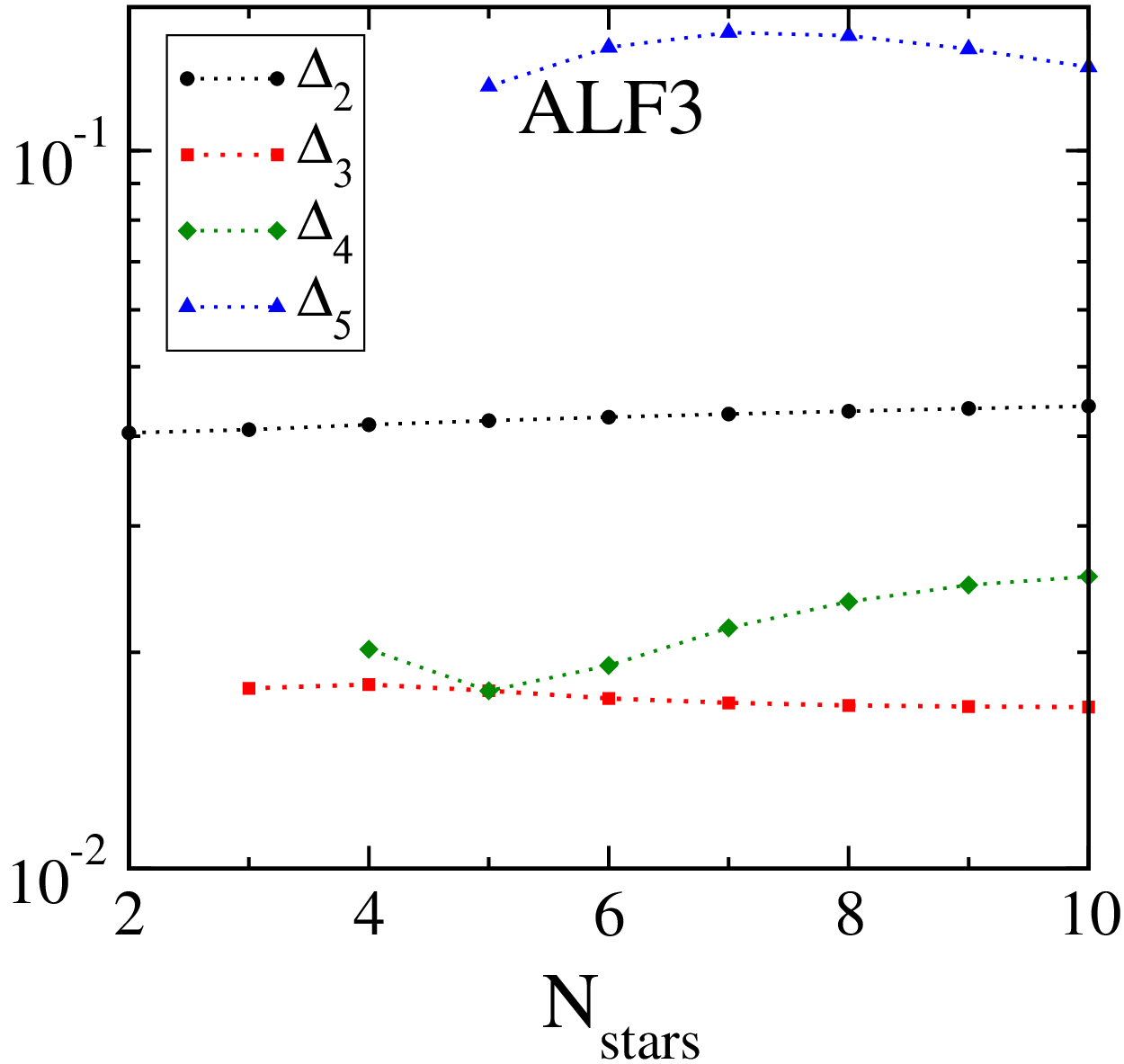}}
\caption{\label{f:alf3_delta_conv} 
Equation of state errors $\Delta_{N_{\gamma_k}}^{MR}$ as functions
of the number of mass-radii data points, $N_\mathrm{stars}$,
used to fix the spectral parameters $\gamma_k$ in
an $N_{\gamma_k}$ parameter spectral expansion.  These results
use mass-radius data computed with the ALF3 equation
of state.}
\end{figure}

The other cases, ALF1 and ALF3, that we have studied in more detail
are more problematic.  The results for the ALF3 case are shown in
Fig.~\ref{f:alf3_delta_conv}, while those for the ALF1 case (not
shown) are similar.  These cases show no sign of the Runge phenomenon,
yet the higher order errors $\Delta_5^{MR}$ (and $\Delta_4^{MR}$ in the ALF1
case) are much larger than the lower order errors $\Delta_2^{MR}$ and
$\Delta_3^{MR}$.  We do not know exactly what is causing this problem in
these cases.  One possibility is that our method for finding the
minimum of $\chi^2(\gamma_k,h_c^i)$ fails for some reason in these
cases for larger values of $N_{\gamma_k}$.  Another possibility is
that these equations of state require more terms in their spectral
expansions before they become truly convergent.  All we can say at
this point is that the spectral representations for these anomalous
cases appear to be more reliable for solutions with smaller numbers of
spectral parameters, i.e., the $N_{\gamma_k}=2$ and $N_{\gamma_k}=3$
cases, than they do for the solutions with larger numbers of
parameters.
%

%%%%%%%%%%%%%%%%%%%%%%%%%%%%%%%%%%%%%%%%%%%%%%%%%%%%%%%%%%%%%%%%%%%%%%%%%%%%%%%
%%%%%%%%%%%%%%%%%%%%%%%%%%%%%%%%%%%%%%%%%%%%%%%%%%%%%%%%%%%%%%%%%%%%%%%%%%%%%%%
\section{Tidal Deformability}
\label{s:TidalDeformability}
%%%%%%%%%%%%%%%%%%%%%%%%%%%%%%%%%%%%%%%%%%%%%%%%%%%%%%%%%%%%%%%%%%%%%%%%%%%%%%%

When a star in a binary system interacts with the tidal field of its
companion, it is deformed by an amount that depends on the internal
structure of that star and hence the equation of state of the material
from which it is made.  These tidal deformations can significantly
effect the phase evolutions of the last parts of the orbits of compact
binary systems, so the gravitational waves emitted by such systems
will contain the imprints of those tidal
interactions~\cite{Kochanek1992, Lai1994, Vallisneri00, Faber2002,
  Will-Mora:2004, Berti2008}.  Accurate observations of the
gravitational waves from neutron-star binary systems will make it
possible therefore to measure the tidal properties of these stars.  A
number of studies~\cite{Hinderer2008a, Hinderer2008, Hinderer2009,
  Read:2009yp, Hinderer2010, Lackey2012, Damour2012, Bernuzzi2012,
  Read2013, Lackey2013, Pozzo2013, Maselli2013} have shown that the
macroscopic neutron-star observable best determined by such
gravitational-wave measurements are the masses $M$ and the tidal
deformabilities $\lambda$.  This section explores the question, How
well can the neutron-star equation of state be determined from
accurate measurements of $M$ and $\lambda$?

The tidal deformability $\lambda$ of a star is defined as the
proportionality factor in the relationship between the tidal field
from a star's companion, ${\cal E}_{ij}$, and the star's quadrupole
moment, $Q_{ij}$, induced by that tidal interaction: $Q_{ij}=-\lambda
{\cal E}_{ij}$. This tidal deformability $\lambda$ is related to the
tidal Love number $k_2$ by $\lambda=2k_2R^5/3$, and to the
dimensionless tidal deformability $\Lambda$: $\Lambda=\lambda/M^5=(2
k_2/3)(R/M)^5$.  Some studies~\cite{Lackey2012, Lackey2013} suggest
that the dimensionless tidal deformability $\Lambda$ can be determined
somewhat more accurately by gravitational wave observations than
$\lambda$, so we use $\Lambda$ in our analysis of this version of the
inverse stellar structure problem.  The equations needed to compute
$\Lambda$ (or equivalently $\lambda$ or $k_2$) for relativistic
neutron stars were first derived by Hinderer~\cite{Hinderer2008,
  Hinderer2009}.  Appendix~\ref{s:AppendixC} presents a more efficient
way to compute $\Lambda(h_c,\gamma_k)$, as well as its derivatives
with respect to the parameters $\gamma_k$ and $h_c$ for the enthalpy
based representations of the parametric equations of state used in our
solution of the inverse stellar structure problem:
$\partial\Lambda/\partial\gamma_k$ and $\partial\Lambda/\partial h_c$.
These derivatives are used by the Levenberg-Marquardt algorithm as
part of our method of finding the global minimum of
$\chi^2(\gamma_k,h_c^i)$.

The spectral approach to the solution of the inverse stellar structure
problem described in Sec.~\ref{s:Introduction} does not depend very
strongly on which macroscopic observables are used.  It is
straightforward to replace the data for observed masses $M_i$ and
radii $R_i$, with those for observed masses $M_i$ and tidal
deformabilities $\Lambda_i$.  The corresponding model observables,
$M(h_c^i,\gamma_k)$ and $\Lambda(h_c^i,\gamma_k)$, are evaluated using
our parametrized equations of state, $\epsilon(h,\gamma_k)$ and
$p(h,\gamma_k)$ with the methods described in
Appendix~\ref{s:AppendixC}.  The equation of state parameters
$\gamma_k$ (and the central enthalpy parameters $h_c^i$) are then
fixed by minimizing the quantity $\chi(\gamma_k,h_c^i)$ that measures
the differences between the observed data and the model observables:
\begin{eqnarray}
\chi^2(\gamma_k,h_c^i)&=&
\frac{1}{N_\mathrm{stars}}\sum_{i=1}^{N_\mathrm{stars}}\left\{
\left[\log\left(\frac{M(h_c^i,\gamma_k)}{M_i}\right)\right]^2\right.\nonumber\\
&&\qquad\qquad\quad
+\left.\left[\log\left(\frac{\Lambda(h_c^i,\gamma_k)}
{\Lambda_i}\right)\right]^2\right\}.
\qquad
\label{e:ChiSquareLMDef}
\end{eqnarray}

We have tested the spectral approach to the relativistic inverse
stellar structure problem (with the improvements described in
Sec.~\ref{s:ImprovingMethod}) using the masses and tidal
deformabilities as observables.  The mock observational data for the
masses and tidal deformabilities used in these tests are based on the
same selections of stellar models computed with the same 34
theoretical high-density neutron-star equations of state used in the
tests described in Sec.~\ref{s:ImprovingMethod}.  The results of these
tests are summarized in Table~\ref{t:TableII}.  For each equation of
state the inverse stellar structure problem has been solved by fitting
$N_{\gamma_k}$ different spectral parameters to mock data sets
containing $N=N_\mathrm{stars}=N_{\gamma_k}$ pairs of mass $M_i$ and
tidal deformability $\Lambda_i$ data.  The minimum value of the
fitting function $\chi_{N}$ is given for each of these solutions in
Table~\ref{t:TableII}.  Two additional quantities,
$\Delta^{M\Lambda}_{N}$ and $\Upsilon^{M\Lambda}_{N}$ are also
included in Table~\ref{t:TableII} that measure how accurately the $N$
parameter spectral equation of state agrees with the original used to
compute the mock mass and tidal deformability observables.  These
equation of state error measures, $\Delta^{M\Lambda}_{N}$ and
$\Upsilon^{M\Lambda}_{N}$, are defined exactly as they were for the
spectral equations of state computed from mass-radius data in
Eqs.~(\ref{e:DeltaMR_Def}) and (\ref{e:UpsilonMR_Def}).
\begin{table*}[!htb]
\begin{center}
\caption{Accuracies of the neutron-star equations of state obtained by
  solving the inverse stellar structure problem.
  $\Delta_N^{M\Lambda}$ measures the average fractional error of the
  equation of state obtained by fitting to $N$ different
  $[M_i,\Lambda_i]$ data pairs.  The parameter $\Upsilon_N^{M\Lambda}$
  measures the ratio of $\Delta_N^{M\Lambda}$ to the accuracy of the
  optimal $N$-parameter spectral fit to each equation of state.  The
  parameter $\chi_N$ measures the accuracy with which the model masses
  $M(h_c^i,\gamma_k)$ and tidal deformability
  $\Lambda(h_c^i,\gamma_k)$ produced by the approximate spectral
  equation of state match the exact $M_i$ and $\Lambda_i$ data.
\label{t:TableII}}
\medskip
\begin{tabular}{|l|cccc|cccc|cccc|}
\hline\hline
    EOS   &$\Delta_{2}^{M\Lambda}$ &$\Delta_{3}^{M\Lambda}$ &$\Delta_{4}^{M\Lambda}$ 
     &$\Delta_{5}^{M\Lambda}$ &$\Upsilon_2^{M\Lambda}$ &$\Upsilon_3^{M\Lambda}$ 
     &$\Upsilon_4^{M\Lambda}$ &$\Upsilon_5^{M\Lambda}$ 
     &$\chi_2$ &$\chi_3$ &$\chi_4$ &$\chi_5$ \\
\hline
    PAL6 &  0.0034 &  0.0019 &  0.0008 &  0.0003 &  1.05 &  1.19 &  1.53 &  2.20 & $ 1.1\times 10^{-11}$ & $ 2.3\times 10^{-11}$ & $ 3.8\times 10^{-11}$ & $ 2.1\times 10^{-11}$ \\
     SLy &  0.0097 &  0.0041 &  0.0024 &  0.0013 &  1.07 &  1.16 &  1.41 &  1.94 & $ 7.5\times 10^{-12}$ & $ 1.3\times 10^{-11}$ & $ 6.1\times 10^{-12}$ & $ 2.1\times 10^{-11}$ \\
    APR1 &  0.0809 &  0.0491 &  0.0384 &  0.0199 &  1.14 &  1.48 &  2.06 &  2.43 & $ 1.7\times 10^{-11}$ & $ 1.7\times 10^{-11}$ & $ 1.8\times 10^{-11}$ & $ 3.3\times 10^{-11}$ \\
    APR2 &  0.0333 &  0.0191 &  0.0111 &  0.0082 &  1.08 &  1.37 &  1.75 &  2.41 & $ 7.2\times 10^{-12}$ & $ 2.3\times 10^{-11}$ & $ 1.8\times 10^{-11}$ & $ 1.6\times 10^{-11}$ \\
    APR3 &  0.0254 &  0.0067 &  0.0035 &  0.0026 &  1.01 &  1.23 &  1.44 &  1.75 & $ 6.8\times 10^{-12}$ & $ 3.4\times 10^{-12}$ & $ 7.0\times 10^{-12}$ & $ 7.0\times 10^{-12}$ \\
    APR4 &  0.0254 &  0.0037 &  0.0021 &  0.0018 &  1.02 &  1.25 &  1.56 &  1.33 & $ 3.8\times 10^{-12}$ & $ 5.6\times 10^{-12}$ & $ 1.7\times 10^{-11}$ & $ 1.6\times 10^{-11}$ \\
     FPS &  0.0046 &  0.0069 &  0.0137 &  0.0076 &  1.03 &  1.63 &  3.60 &  4.18 & $ 1.1\times 10^{-11}$ & $ 3.6\times 10^{-11}$ & $ 7.4\times 10^{-12}$ & $ 2.1\times 10^{-11}$ \\
    WFF1 &  0.0599 &  0.0212 &  0.0340 &  0.0290 &  1.13 &  1.99 &  4.93 &  4.43 & $ 7.3\times 10^{-12}$ & $ 1.3\times 10^{-11}$ & $ 4.5\times 10^{-12}$ & $ 2.2\times 10^{-11}$ \\
    WFF2 &  0.0294 &  0.0172 &  0.0088 &  0.0055 &  1.08 &  1.43 &  1.23 &  1.45 & $ 3.0\times 10^{-12}$ & $ 1.3\times 10^{-11}$ & $ 1.2\times 10^{-11}$ & $ 2.0\times 10^{-11}$ \\
    WFF3 &  0.0141 &  0.0192 &  0.0190 &  0.0124 &  1.27 &  1.86 &  3.19 &  2.24 & $ 5.7\times 10^{-12}$ & $ 7.6\times 10^{-12}$ & $ 3.7\times 10^{-12}$ & $ 8.0\times 10^{-11}$ \\
    BBB2 &  0.0344 &  0.0368 &  0.0357 &  0.0143 &  1.04 &  1.28 &  1.64 &  1.55 & $ 6.8\times 10^{-12}$ & $ 6.0\times 10^{-12}$ & $ 1.9\times 10^{-11}$ & $ 2.5\times 10^{-11}$ \\
  BPAL12 &  0.0184 &  0.0118 &  0.0076 &  0.0090 &  1.07 &  1.19 &  1.54 &  4.04 & $ 9.3\times 10^{-12}$ & $ 2.2\times 10^{-11}$ & $ 1.3\times 10^{-11}$ & $ 9.8\times 10^{-11}$ \\
     ENG &  0.0219 &  0.0243 &  0.0207 &  0.0520 &  1.08 &  1.31 &  1.40 &  4.62 & $ 4.0\times 10^{-12}$ & $ 4.1\times 10^{-12}$ & $ 2.2\times 10^{-11}$ & $ 1.7\times 10^{-11}$ \\
    MPA1 &  0.0301 &  0.0043 &  0.0061 &  0.0081 &  1.17 &  1.33 &  1.98 &  3.58 & $ 1.4\times 10^{-11}$ & $ 1.3\times 10^{-11}$ & $ 1.5\times 10^{-11}$ & $ 1.6\times 10^{-11}$ \\
     MS1 &  0.0465 &  0.0141 &  0.0129 &  0.0008 &  1.62 &  2.49 &  3.56 &  2.44 & $ 1.7\times 10^{-11}$ & $ 1.7\times 10^{-11}$ & $ 9.8\times 10^{-12}$ & $ 1.9\times 10^{-11}$ \\
     MS2 &  0.0155 &  0.0042 &  0.0009 &  0.0005 &  1.32 &  1.80 &  2.18 &  3.20 & $ 1.6\times 10^{-13}$ & $ 4.2\times 10^{-13}$ & $ 5.6\times 10^{-13}$ & $ 5.7\times 10^{-13}$ \\
    MS1B &  0.0304 &  0.0135 &  0.0084 &  0.0014 &  1.52 &  2.10 &  2.82 &  5.08 & $ 8.0\times 10^{-12}$ & $ 5.1\times 10^{-12}$ & $ 1.4\times 10^{-11}$ & $ 1.5\times 10^{-11}$ \\
      PS &  0.1044 &  0.0740 &  0.1120 &  0.0439 &  1.66 &  2.46 &  3.73 &  2.62 & $ 3.7\times 10^{-12}$ & $ 1.4\times 10^{-11}$ & $ 2.9\times 10^{-11}$ & $ 8.0\times 10^{-5}$ \\
     GS1 &  0.1018 &  0.0648 &  0.0386 &  0.0493 &  1.14 &  1.68 &  1.02 &  1.97 & $ 3.0\times 10^{-12}$ & $ 1.3\times 10^{-12}$ & $ 2.6\times 10^{-12}$ & $ 9.5\times 10^{-12}$ \\
     GS2 &  0.0909 &  0.0855 &  0.1164 &  0.0537 &  1.50 &  1.95 &  2.67 &  1.70 & $ 2.7\times 10^{-12}$ & $ 4.3\times 10^{-12}$ & $ 1.6\times 10^{-11}$ & $ 1.9\times 10^{-11}$ \\
  BGN1H1 &  0.1356 &  0.1652 &  0.1445 &  0.1363 &  1.55 &  3.30 &  3.26 &  3.89 & $ 1.6\times 10^{-11}$ & $ 1.7\times 10^{-11}$ & $ 3.5\times 10^{-11}$ & $ 4.6\times 10^{-11}$ \\
    GNH3 &  0.0182 &  0.0171 &  0.0397 &  0.0216 &  1.32 &  1.80 &  4.82 &  3.70 & $ 7.2\times 10^{-12}$ & $ 5.7\times 10^{-12}$ & $ 6.3\times 10^{-12}$ & $ 4.3\times 10^{-11}$ \\
      H1 &  0.0309 &  0.0154 &  0.0124 &  0.0107 &  1.51 &  1.23 &  1.45 &  1.49 & $ 2.5\times 10^{-11}$ & $ 3.6\times 10^{-11}$ & $ 2.1\times 10^{-11}$ & $ 3.6\times 10^{-11}$ \\
      H2 &  0.0226 &  0.0265 &  0.0153 &  0.0263 &  1.27 &  1.90 &  2.22 &  3.83 & $ 1.6\times 10^{-11}$ & $ 1.1\times 10^{-11}$ & $ 1.5\times 10^{-11}$ & $ 1.5\times 10^{-11}$ \\
      H3 &  0.0151 &  0.0186 &  0.0177 &  0.0118 &  1.18 &  1.66 &  2.09 &  1.70 & $ 2.0\times 10^{-11}$ & $ 1.4\times 10^{-11}$ & $ 1.2\times 10^{-11}$ & $ 4.2\times 10^{-11}$ \\
      H4 &  0.0119 &  0.0256 &  0.0211 &  0.0141 &  1.15 &  2.57 &  3.11 &  2.09 & $ 2.1\times 10^{-11}$ & $ 1.2\times 10^{-11}$ & $ 1.6\times 10^{-11}$ & $ 2.1\times 10^{-11}$ \\
      H5 &  0.0141 &  0.0293 &  0.0145 &  0.0221 &  1.03 &  2.19 &  2.46 &  4.48 & $ 7.5\times 10^{-12}$ & $ 1.9\times 10^{-11}$ & $ 9.7\times 10^{-12}$ & $ 1.4\times 10^{-11}$ \\
      H6 &  0.0160 &  0.0144 &  0.0204 &  0.0160 &  1.16 &  1.05 &  1.56 &  1.40 & $ 1.1\times 10^{-11}$ & $ 1.2\times 10^{-11}$ & $ 1.2\times 10^{-11}$ & $ 1.3\times 10^{-11}$ \\
      H7 &  0.0142 &  0.0205 &  0.0136 &  0.0170 &  1.16 &  1.83 &  2.38 &  3.00 & $ 8.7\times 10^{-12}$ & $ 1.1\times 10^{-11}$ & $ 2.0\times 10^{-11}$ & $ 2.3\times 10^{-11}$ \\
    PCL2 &  0.0378 &  0.0154 &  0.0103 &  0.0288 &  1.37 &  1.18 &  1.07 &  3.52 & $ 1.2\times 10^{-11}$ & $ 2.4\times 10^{-11}$ & $ 2.2\times 10^{-11}$ & $ 4.3\times 10^{-11}$ \\
    ALF1 &  0.0795 &  0.0704 &  0.1427 &  0.1225 &  1.08 &  1.47 &  3.55 &  4.31 & $ 3.8\times 10^{-11}$ & $ 8.9\times 10^{-12}$ & $ 3.8\times 10^{-11}$ & $ 3.3\times 10^{-11}$ \\
    ALF2 &  0.0725 &  0.0630 &  0.0479 &  0.0225 &  1.04 &  1.28 &  1.73 &  1.26 & $ 1.3\times 10^{-11}$ & $ 1.1\times 10^{-11}$ & $ 1.6\times 10^{-11}$ & $ 2.3\times 10^{-11}$ \\
    ALF3 &  0.0408 &  0.0203 &  0.0200 &  0.1566 &  1.05 &  1.36 &  1.42 & 11.64 & $ 1.1\times 10^{-11}$ & $ 1.2\times 10^{-11}$ & $ 2.8\times 10^{-11}$ & $ 9.2\times 10^{-11}$ \\
    ALF4 &  0.0793 &  0.0193 &  0.0213 &  0.0600 &  1.12 &  1.43 &  2.14 &  6.33 & $ 7.9\times 10^{-12}$ & $ 1.3\times 10^{-11}$ & $ 1.5\times 10^{-11}$ & $ 3.0\times 10^{-11}$ \\
\hline
Averages &  0.0403 &  0.0295 &  0.0304 &  0.0291 &  1.23 &  1.69 &  2.40 &  3.30 & &&&\\
\hline\hline
 \end{tabular}
\end{center}
\end{table*}

The results for the $M\Lambda$ case shown in Table~\ref{t:TableII} are
very similar, both quantitatively and qualitatively, to those from the
$MR$ case shown in Table~\ref{t:TableI}.  All of the $\chi_N$ in
Table~\ref{t:TableII} meet our convergence criterion
$\chi_N<10^{-10}$, except the $N_{\gamma_k}=5$ case of the PS equation
of state.  This is the same exceptional case as in
Table~\ref{t:TableI}, suggesting there is some pathology in this
particular equation of state that keeps our code from finding accurate
reproducible solutions to the standard stellar structure problem.
Similar problems were eliminated when we corrected the non-monotonicity
problems in some of the equations of state, as described in
Sec.~\ref{s:ImprovingMethod}.  Unfortunately, we have not been able to
identify any similar problem with the PS equation of state.

The parameters $\Delta_N^{M\Lambda}$ in Table~\ref{t:TableII} that
quantify the errors in the spectral equations of state for the
$M\Lambda$ case are very similar to those found using using $MR$ data
in Table~\ref{t:TableI} .  The averages of these quantities (over the
34 different theoretical equations of state) in these tests are
$\Delta_2^{M\Lambda}=0.040$, $\Delta_3^{M\Lambda}=0.029$,
$\Delta_4^{M\Lambda}=0.028$, $\Delta_5^{M\Lambda}=0.024$, while those
found in the $MR$ case were $\Delta_2^{MR}=0.040$,
$\Delta_3^{MR}=0.030$, $\Delta_4^{MR}=0.030$, $\Delta_5^{MR}=0.029$.
The errors in the $M\Lambda$ cases with $N_{\gamma_k}=2$ and
$N_{\gamma_k}=3$ are almost identical to those from the analogous $MR$
cases.  But the errors in the cases with $N_{\gamma_k}=4$ and
$N_{\gamma_k}=5$ are slightly larger.  We don't know exactly why.  We
note that the $M\Lambda$ cases with poorest convergence properties are
the same ones that show poor convergence using $MR$ data.  This
suggests that this anomalous behavior may be caused by some
pathological feature of these particular equations of state, rather
than some general failure of the method itself.

%%%%%%%%%%%%%%%%%%%%%%%%%%%%%%%%%%%%%%%%%%%%%%%%%%%%%%%%%%%%%%%%%%%%%%%%%%%%%%%
%%%%%%%%%%%%%%%%%%%%%%%%%%%%%%%%%%%%%%%%%%%%%%%%%%%%%%%%%%%%%%%%%%%%%%%%%%%%%%%
\section{Discussion}
\label{s:Discussion}
%%%%%%%%%%%%%%%%%%%%%%%%%%%%%%%%%%%%%%%%%%%%%%%%%%%%%%%%%%%%%%%%%%%%%%%%%%%%%%%

In summary, we have improved our method of solving the relativistic
inverse stellar structure problem using faithful spectral expansions
of the unknown high density part of the equation of state.  This
method is based on minimizing a function $\chi$ that measures the
differences between a given set of observables, e.g. $[M_i,R_i]$, and
model values of these observables, e.g. $M(h_c^i,\gamma_k)$ and
$R(h_c^i,\gamma_k)$.  Our improved methods described in
Sec.~\ref{s:ImprovingMethod} are much better at finding the global
minimum of this complicated non-linear function $\chi$ of the model
parameters $\gamma_k$ and $h_c^i$.  The numerical tests of our
improved method, described in Sec.~\ref{s:ImprovingMethod},
consistently give much smaller values of $\chi$ than those in the
tests of our original method~\cite{Lindblom12}.  We have also expanded
our new method in Sec.~\ref{s:TidalDeformability} to solve the inverse
stellar structure problem using the mass and tidal deformability of a
star as the observables: $[M_i,\Lambda_i]$.  To do this we have
developed (in Appendix~\ref{s:AppendixC}) more efficient and accurate
ways to evaluate the tidal deformability $\Lambda(h_c,\gamma_k)$ and
its derivatives with respect to $h_c$ and $\gamma_k$.  The tests of
our solution to the $[M_i,\Lambda_i]$ version of the inverse stellar
structure problem show that accurate measurements of $[M_i,\Lambda_i]$
data can determine the neutron-star equation of state about as
accurately as it could using the same number of accurate $[M_i,R_i]$
data.  Using only two $[M_i,R_i]$ or $[M_i,\Lambda_i]$ data points,
this new method can determine the high density part of the
neutron-star equation of state that is present in these stars with
errors (on average) of just a few percent.

Our analysis of the relativistic inverse stellar structure problem,
introduced in Refs.~\cite{Lindblom12,Lindblom1992} and continued here
in Secs.~\ref{s:ImprovingMethod} and \ref{s:TidalDeformability}, has
focused on understanding some of the fundamental mathematical aspects
of this problem.  Is it possible to determine the neutron-star
equation of state exactly from a complete knowledge of the macroscopic
observable properties of these stars, i.e., does this problem have a
unique solution?  Can numerical methods be devised whose approximate
solutions converge to the exact equation of state when a complete
exact knowledge of the macroscopic observables of these stars is
available?  What level of numerical approximation and how many
macroscopic observable data points are needed to achieve reasonable
levels of accuracy for ``realistic'' neutron-star equations of state?
While various observational and data-analysis questions related to
this problem have been studied previously by a number of researchers,
our studies of these fundamental questions are unique (to our
knowledge).

An essential element of any practical robust solution to the inverse
stellar structure problem (in our opinion) is the use of faithful
parametric representations of the equation of state.  These faithful
representations must not exclude any physically possible equation of
state, and conversely no choice of parameters may correspond to a
physically impossible equation of state.  To our knowledge the only
faithful parametric representations of the high density equation of
state discussed in the literature are the piecewise-polytropic
representations of Read, et al.~\cite{Read:2008iy}, and our spectral
representations~\cite{Lindblom10} (which in general are somewhat more
accurate for a given number of parameters than the
piecewise-polytropes).

\"Ozel and collaborators~\cite{Ozel2009, Guver2012a, Guver2013} and
Steiner and collaborators~\cite{Steiner2010, Steiner2012} have used
low-order piecewise-polytropic models of the equation of state to
solve the inverse stellar structure problem using presently available
mass and radius measurements of neutron stars.  Both groups have
studied the accuracy with which the presently available $[M_i,R_i]$
data have been determined observationally.  Both groups have done
careful studies of the effects of these measurement errors on the
accuracy with which the parameters in their high-density equation of
state models are determined in this way.  However, neither group has
considered some of the more fundamental questions like those studied
here, e.g., how accurately their solutions to the inverse stellar
structure problem represent the actual neutron-star equation of state,
or whether their method converges when higher-order parametric
equation of state models are used in the solution.
  
A number of researchers have shown that tidal effects in compact
binary systems can influence the gravitational waveforms they emit in
an equation of state dependent way~\cite{Kochanek1992, Lai1994,
  Vallisneri00, Faber2002, Will-Mora:2004, Berti2008}.  Flanagan and
Hinderer showed that a neutron-star's tidal deformability was the
particular stellar characteristic that determines the leading-order
effect on these gravitational waveforms~\cite{Hinderer2008a}.
Hinderer was the first to derive the equations that determine the
tidal deformability from the structure of a relativistic stellar
model~\cite{Hinderer2008, Hinderer2009}.  Hinderer and collaborators
were the first to explore how the tidal deformability depends on the
equation of state by evaluating it numerically for a number of
theoretical neutron-star equations of state~\cite{Hinderer2010}.  We
have extended this basic formalism for evaluating the tidal
deformability in this paper in two important ways.  First, we derive
(in Appendix~\ref{s:AppendixC}) an expression for the tidal
deformability in terms of a solution to a first-order, rather than a
second-order, differential equation.  Our expression can therefore be
evaluated numerically more accurately and efficiently.  Second, we
derive a set of differential equations whose solutions determine the
variations of the tidal deformability with respect to the equation of
state parameters.  These expressions make it possible to determine
these equation of state parameters from tidal deformability data more
accurately and efficiently.

A number of researchers have studied how the tidal deformability of
neutron stars can be measured from observations of the gravitational
waves emitted by compact binary systems~\cite{Read:2009yp,
  Hinderer2010, Lackey2012, Damour2012, Bernuzzi2012, Lackey2013,
  Pozzo2013, Maselli2013, Read2013}.  These researchers have
constructed post-Newtonian~\cite{Hinderer2010, Pozzo2013,
  Maselli2013}, effective one body~\cite{Damour2012}, and numerical
relativity models~\cite{Read:2009yp, Lackey2012, Bernuzzi2012,
  Read2013, Lackey2013} of the waveforms produced by these systems.
They have also explored in great detail (using a variety of
data-analysis methods) the expected accuracy with which the tidal
deformability should be measured by the next generation of
gravitational wave detectors (advanced LIGO, etc.).  These researchers
have shown, for example, that such measurements are likely to be
accurate enough to distinguish between some of the published
theoretical neutron-star equations of state.  None of these studies,
however, has considered any of the more fundamental questions about
the relativistic inverse stellar structure problem that we consider
here.  They have not proposed a method for determining the equation of
state itself from these gravitational wave measurements, nor have they
estimated how accurately it could be determined.  Our study presented
in Sec.~\ref{s:TidalDeformability} of this paper is therefore unique
(to our knowledge) in its exploration of some of the fundamental
questions associated with the mass and tidal deformability version of
the inverse stellar structure problem.

The spectral approach to the solution of the inverse stellar structure
problem introduced in Ref.~\cite{Lindblom12} and improved and extended
in Secs.~\ref{s:ImprovingMethod} and \ref{s:TidalDeformability} of
this paper has been shown to be quite effective in determining the
high-density neutron-star equation of state using high-accuracy
measurements of the mass and radius (or the mass and tidal
deformability) of just two or three neutron stars.  However, many
basic questions remain unanswered.  The equations of state produced by
our current implementation of the spectral approach do not converge to
the exact equation of state in a few cases as the number of
observational data points is increased.  At the present time we do not
understand the reason for this.  More study of the mathematical
properties of the inverse stellar structure problem is therefore
needed to resolve these remaining questions.

Our studies of the inverse stellar structure problem have also assumed
that the observational data were ideal.  Additional research is
therefore needed to explore the robustness of our approach before it
can be used as a practical tool for analyzing observational data.  How
do the errors in the approximate spectral equations of state change
when more realistic $[M_i,R_i]$ or $[M_i,\Lambda_i]$ data are used?
The data used in our tests were idealized in two important ways.
First, the mock $[M_i,R_i]$ or $[M_i,\Lambda_i]$ data were supplied
with very high precision.  Real astrophysical measurements of these
quantities will have significant errors.  How will measurement errors
influence the accuracy of the equation of state that is constructed by
these techniques?  Second, the mock $[M_i,R_i]$ or $[M_i,\Lambda_i]$
data used in our tests were chosen to cover uniformly the
astrophysically relevant range of neutron-star masses.  Real
astrophysical measurements will not be distributed in such a complete
and orderly way.  How will the accuracy of the implied equation of
state be affected by different, presumably less ideal, data
distributions?  In particular, how does the accuracy of the
highest-density part of the equation of state depend on the mass of
the most massive neutron-star for which observational data are
available?

%%%%%%%%%%%%%%%%%%%%%%%%%%%%%%%%%%%%%%%%%%%%%%%%%%%%%%%%%%%%%%%%%%%%%%%%%%%%%%%
%Acknowledgments
%%%%%%%%%%%%%%%%%%%%%%%%%%%%%%%%%%%%%%%%%%%%%%%%%%%%%%%%%%%%%%%%%%%%%%%%%%%%%%%

\acknowledgments We thank John Friedman, Tanja Hinderer, Benjamin
Lackey, and Manuel Tiglio for helpful discussions concerning this
research.  A portion of this research was carried out during the time
L.L. was a visitor at the Leonard E. Parker Center for Gravitation,
Cosmology and Astrophysics, University of Wisconsin at Milwaukee.
This research was supported in part by a grant from the Sherman
Fairchild Foundation and by NSF grants PHY1005655 and DMS1065438.

%%%%%%%%%%%%%%%%%%%%%%%%%%%%%%%%%%%%%%%%%%%%%%%%%%%%%%%%%%%%%%%%%%%%%%%%%%%%%%%
% Appendix
%%%%%%%%%%%%%%%%%%%%%%%%%%%%%%%%%%%%%%%%%%%%%%%%%%%%%%%%%%%%%%%%%%%%%%%%%%%%%%%

\appendix
 
%%%%%%%%%%%%%%%%%%%%%%%%%%%%%%%%%%%%%%%%%%%%%%%%%%%%%%%%%%%%%%%%%%%%%%%%%%%%%%%
\section{Estimating $h_\infty(\gamma_k)$}
\label{s:AppendixA}
%%%%%%%%%%%%%%%%%%%%%%%%%%%%%%%%%%%%%%%%%%%%%%%%%%%%%%%%%%%%%%%%%%%%%%%%%%%%%%%

The parametric representation of the equation of state used in our
analysis, $\epsilon=\epsilon(h,\gamma_k)$ and $p=p(h,\gamma_k)$, is
constructed from a spectral expansion of the adiabatic
index $\Gamma(h)$ of the material~\cite{Lindblom10}:
\begin{eqnarray}
\Gamma(h) &\equiv& \frac{\epsilon + p}{p}\frac{dp}{d\epsilon}
=\frac{\epsilon + p}{p}\frac{dp}{dh}
\left(\frac{d\epsilon}{dh}\right)^{-1},
 \label{e:GammaDef}\\  
&=& \exp\left\{\sum_k \gamma_k 
\left[\log\left(\frac{h}{h_0}\right)\right]^k\right\},
\label{e:Gamma}
\end{eqnarray}
where $h_0$ is the lower bound on the enthalpy, $h_0\leq h$, in the
domain where the spectral expansion is to be used.  This is a standard
spectral expansion of the function $\log \Gamma(h)$ in which the
$[\log(h/h_0)]^k$ are the spectral basis functions and the 
$\gamma_k$ are the spectral expansion coefficients (or parameters).

The equation of state functions $p(h,\gamma_k)$ and
$\epsilon(h,\gamma_k)$ are obtained from $\Gamma(h,\gamma_k)$ by integrating
the system of ordinary differential equations,
\begin{eqnarray}
\frac{dp}{dh} &=& \epsilon + p,\label{e:dpdh}
\\
\frac{d\epsilon}{dh} &=& \frac{(\epsilon + p)^2}{p  \Gamma(h)},
\label{e:depsilondh}
\end{eqnarray}
that follow from the definitions of $h$ and $\Gamma$ in
Eqs.~(\ref{e:EnthalpyDef}) and (\ref{e:GammaDef}).  The general
solution to these equations can be reduced to quadratures:
\begin{eqnarray}
p(h)&=&p_0 \exp\left[\int_{h_0}^h \frac{e^{h'}dh'}{\mu(h')}
\right],\label{e:PressueH}\\
\epsilon(h)&=&p(h)  \frac{e^h -\mu(h)}{\mu(h)},
\label{e:EnthalpyH}
\end{eqnarray}
where $\mu(h)$ is defined as.
\begin{eqnarray}
\mu(h) = \frac{p_0\, e^{h_0}}{\epsilon_0  + p_0} 
+ \int_{h_0}^h \frac{\Gamma(h')-1}{\Gamma(h')} e^{h'}dh'.
\label{e:TildeMuDef}
\end{eqnarray}
The constants $p_0$ and $\epsilon_0$ are defined by $p_0=p(h_0)$ and
$\epsilon_0=\epsilon(h_0)$ respectively. 

Equations~(\ref{e:PressueH})--(\ref{e:TildeMuDef}) show that
$\epsilon(h)$ and $p(h)$ are finite (for $h_0\leq h < \infty$) unless
there exists an $h=h_\infty$ where $\mu(h_\infty)=0$.  The problem of
finding $h_\infty$ is reduced therefore to the problem of finding the
first zero of $\mu(h)$ above $h_0$.  It is not necessary for our
purposes to know the exact value of $h_\infty$.  Rather a firm
estimate $h_{\max}<h_\infty$ that is beyond the range of $h$ occurring
in neutron stars is all that is needed.

Equation~(\ref{e:TildeMuDef}) shows that $\mu(h_0)>0$ and that
$\mu(h)$ is monotonically increasing unless $\Gamma(h)<1$.  The first
step in finding a useful estimate $h_{\max}$ is to evaluate
$\Gamma(h)$ (which can be done very efficiently) on a mesh of points
covering the range $h_0\leq h \leq h_0 e^5$.  If $\Gamma(h)\geq 1$
throughout this range, then we simply set $h_{\max}=h_0e^5$.  The
upper limit of this range needs to be larger than any value of $h$
that is likely to occur within a neutron star.  For the cases we have
studied the value $h_0e^5$ is a factor of 4 or 5 larger than any $h$
we have seen in a neutron-star model, but its value could (and should)
be adjusted upward as needed.  If one of the mesh points, $h_n$, is
found where $\Gamma(h_n)<1$, then we evaluate $\mu(h)$ on a second
mesh of points that covers the range $h_n\leq h \leq h_0e^5$.  If
$\mu(h)$ is positive throughout this range, then we again set
$h_{\max}=h_0e^5$.  If $\mu(h)$ is found to become negative somewhere
in this range then we use standard numerical root finding methods to
determine the location of $h_\infty$ where $\mu(h_\infty)=0$.  In this
case we set $h_{\max}=h_\infty$.

%%%%%%%%%%%%%%%%%%%%%%%%%%%%%%%%%%%%%%%%%%%%%%%%%%%%%%%%%%%%%%%%%%%%%%%%%%%%%%%
\section{Interpolating and Extrapolating Equation of State Tables}
\label{s:AppendixB}
%%%%%%%%%%%%%%%%%%%%%%%%%%%%%%%%%%%%%%%%%%%%%%%%%%%%%%%%%%%%%%%%%%%%%%%%%%%%%%%

This appendix describes the method for interpolating between table
entries for the exact equation of states used in the tests described
here.  This change was motivated by our need to find the tidal
deformabilities $\Lambda$ of stellar models with these equations of
state.  The equations that determine $\Lambda$ depend on the adiabatic
index of the material.  In our original work the equation of state
below the first table entry was assumed to have uniform density, and
therefore infinite adiabatic index.  This choice made it difficult
therefore to evaluate $\Lambda$.  Consequently the method used here to
extrapolate below the lowest table entries has been changed.  For
clarity, this appendix provides a complete description of the
interpolation methods used in this paper.  We assume that the
exact equation of state is represented as a table of energy densities
$\epsilon_i$ and corresponding pressures $p_i$ for $i=1,...,N$.  For
our purposes here we will convert these to an equation of state of the
form $\epsilon=\epsilon(h)$ and $p=p(h)$ in the following way.  We do
this by assuming that the exact equation of state is obtained for
values intermediate between those given in the table, $\epsilon_i \leq
\epsilon\leq \epsilon_{i+1}$, by the interpolation formula:
\begin{eqnarray}
\frac{p}{p_i} &=& \left(\frac{\epsilon}{\epsilon_i}\right)^{c_{i+1}},
\label{e:piInterpolate}\\
c_{i+1}&=&\frac{\log(p_{i+1}/p_i)}{\log(\epsilon_{i+1}/\epsilon_i)}.
\end{eqnarray}
For smaller values of the density than the lowest
entry in the table, $\epsilon\leq \epsilon_1$, we
assume,
\begin{eqnarray}
\frac{p}{p_1} &=& \left(\frac{\epsilon}{\epsilon_1}\right)^{5/3},
\label{e:p1Interpolate}
\end{eqnarray}
and for larger values of the density than the highest entry,
$\epsilon\geq\epsilon_{N}$, we assume,
\begin{eqnarray}
\frac{p}{p_{N}} 
&=& \left(\frac{\epsilon}{\epsilon_{N}}\right)^{c_{N}}.
\label{e:pNInterpolate}
\end{eqnarray}
The low density extrapolation given in Eq.~(\ref{e:p1Interpolate})
assumes that the equation of state is that of a low temperature
non-relativistic Fermi gas with adiabatic index 5/3, while the high
density extrapolation given in Eq.~(\ref{e:pNInterpolate}) just
extends the tabulated portion of the equation of state smoothly.

Given this prescription for interpolation, it is straightforward to
show that the values of the enthalpy
\begin{eqnarray}
h(p)=\int_0^p\frac{dp'}{\epsilon(p')+p'},
\label{e:AppendixEnthalpy}
\end{eqnarray}
are given at the table entry values $h_i=h(p_i)$, by
\begin{eqnarray}
h_1&=& \frac{5}{2}\log\left(\frac{\epsilon_1+p_1}{\epsilon_1}\right),\\
h_{i+1}&=& h_i+\frac{c_{i+1}}{c_{i+1}-1}
\log\left[\frac{\epsilon_i(\epsilon_{i+1}+p_{i+1})}
{\epsilon_{i+1}(\epsilon_i+p_i)}\right].
\end{eqnarray}

The pressure is determined as a function of the enthalpy, by
performing the integral in Eq.~(\ref{e:AppendixEnthalpy}) to give
$h(p)$, and then inverting.  It is slightly more convenient to perform
this inversion to give $\epsilon(h)$, from which it is straightforward
to determine $p(h)$ through Eqs.~(\ref{e:p1Interpolate}) and
(\ref{e:piInterpolate}):
\begin{eqnarray}
\!\!\!\!\!\epsilon(h)&=& \epsilon_1\left\{\frac{\epsilon_1}{p_1}
\left[\exp\left(\frac{2 h}{5}\right)-1\right]
\right\}^{3/2}
\end{eqnarray}
for $h\leq h_1$,
\begin{eqnarray}
&&\!\!\!\!\!\epsilon(h)=\nonumber\\
&&\!\!\!\!\! \epsilon_i\left\{\frac{\epsilon_i+p_i}{p_i}
\exp\left[\frac{c_{i+1}-1}{c_{i+1}} (h-h_i)\right]-\frac{\epsilon_i}{p_i}
\right\}^{1/(c_{i+1}-1)}
\end{eqnarray}
for $h_i\leq h \leq h_{i+1}$, and
\begin{eqnarray}
&&\!\!\!\!\!\epsilon(h)=\nonumber\\
&&\!\!\!\!\! \epsilon_N\left\{\frac{\epsilon_N+p_N}{p_N}
\exp\left[\frac{c_{N}-1}{c_{N}} (h-h_N)\right]-\frac{\epsilon_N}{p_N}
\right\}^{1/(c_{N}-1)},
\end{eqnarray}
for $h\geq h_N$.

%%%%%%%%%%%%%%%%%%%%%%%%%%%%%%%%%%%%%%%%%%%%%%%%%%%%%%%%%%%%%%%%%%%%%%%%%%%%%%%
\section{Computing $\Lambda$ and its Derivatives}
\label{s:AppendixC}
%%%%%%%%%%%%%%%%%%%%%%%%%%%%%%%%%%%%%%%%%%%%%%%%%%%%%%%%%%%%%%%%%%%%%%%%%%%%%%%

A number of studies~\cite{Hinderer2008a, Hinderer2008, Hinderer2009,
  Read:2009yp, Hinderer2010, Lackey2012, Damour2012, Bernuzzi2012,
  Read2013, Lackey2013, Pozzo2013, Maselli2013} have shown that the
mass $M$ and the tidal deformability $\lambda$ are the neutron-star
observables best measured by gravitational wave observations of
neutron-star binary systems, while some studies~\cite{Lackey2012,
  Lackey2013} suggest that the dimensionless tidal deformability
$\Lambda=\lambda/M^5$ can be determined somewhat more accurately than
$\lambda$ itself.  Hinderer~\cite{Hinderer2008, Hinderer2009} derived
the expressions for the tidal deformability $\lambda$, or equivalently
the dimensionless tidal deformability $\Lambda$ of a relativistic
stellar model, in terms of the gravitational compactness $C=M/R$ and a
quantity $Y$ that measures the relativistic quadrupole gravitational
potential induced by the tidal deformation.  Using those expressions,
the dimensionless tidal deformability $\Lambda$ can be expressed in
terms of $C$ and $Y$ in the following way,
\begin{eqnarray}
&&\!\!\!\!\!
\Lambda(C,Y)=
\frac{16}{15 \Xi}(1-2C)^2[2+2C(Y-1)-Y],\qquad
\label{e:DefTidalDeformationAppendix}
\end{eqnarray}
where $\Xi$ is given by
\begin{eqnarray}
&&
\!\!\!\!\!
\Xi(C,Y)=4C^3[13-11Y+C(3Y-2)+2C^2(1+Y)]\nonumber\\
&&\qquad+3(1-2C)^2[2-Y+2C(Y-1)]\log(1-2C)\nonumber\\
&&\qquad+2C[6-3Y+3C(5Y-8)].
\end{eqnarray}
This dimensionless tidal deformability $\Lambda$ is the observable we
use in our study of the inverse stellar structure problem in
Sec.~\ref{s:TidalDeformability} of this paper.

The gravitational compactness $C=M/R$ of a relativistic stellar model
is computed by solving the Oppenheimer-Volkoff equations:
\begin{eqnarray}
\frac{dm}{dr}&=&4\pi r^2\epsilon,
\label{e:OVMass}\\
\frac{dp}{dr}&=&-(\epsilon+p)\frac{m+4\pi r^3 p}{r(r-2m)}.
\label{e:OVPressure}
\end{eqnarray}
The radius of the star, $R$, is the surface where the pressure
vanishes, $p(R)=0$, while the star's total mass, $M$, is
$M=m(R)$.\footnote{ In this paper we use geometrical units in which
  the gravitational constant $G$ and the speed of light $c$ are one:
  $G=c=1$.}  The relativistic quadrupole gravitational potential, $H$,
induced by the tidal interaction is determined by solving the
Regge-Wheeler equation (cf Hinderer~\cite{Hinderer2008,
  Hinderer2009}):
\begin{eqnarray}
0&=&\frac{d^2 H}{dr^2} + \left[\frac{2}{r}
+\frac{2m +4\pi r^3(p-\epsilon)}{r(r-2m)}\right]\frac{dH}{dr}\nonumber\\
&&\qquad+\Biggl\{4\pi r\left[5\epsilon+9p
+\frac{(\epsilon+p)^2}{p\Gamma}\right]\nonumber\\
&&\qquad\qquad -\frac{6}{r}
-\frac{4(m+4\pi r^3 p)^2}{r^2(r-2m)}\Biggr\}\frac{H}{r-2m}.
\qquad
\label{e:ReggeWheeler}
\end{eqnarray}
The potential $Y$ that appears in the expression for the tidal
deformability $\Lambda$, Eq.~(\ref{e:DefTidalDeformationAppendix}), is
defined as $Y=(R/H)(dH/dr)$ evaluated at the surface of the star.
Since $H$ itself does not enter the expression for $\Lambda$,
it is more efficient to transform Eq.~(\ref{e:ReggeWheeler}) into
a form that determines only the part of the potential that is needed:
\begin{eqnarray}
y = \frac{r}{H}\frac{dH}{dr}.
\end{eqnarray}
The resulting first-order equation for $y$ is given by,
\begin{eqnarray}
\frac{dy}{dr}&=&  -\frac{y^2}{r} - \frac{r+4\pi r^3 (p-\epsilon)}{r(r-2m)}y
+\frac{4(m+4\pi r^3 p)^2}{r(r-2m)^2}\quad\nonumber\\
&&
+\frac{6}{r-2m}
-\frac{4\pi r^2}{r-2m}\left[5\epsilon+9p
+\frac{(\epsilon+p)^2}{p\Gamma}\right].
\label{e:y_equation_def}
\end{eqnarray}
The potential $Y$ that appears in the expression for $\Lambda$ is just
the surface value of the potential $y$ determined by solving
Eq.~(\ref{e:y_equation_def}): $Y=y(R)$.  The solutions to
Eqs.~(\ref{e:OVMass}), (\ref{e:OVPressure}), and
(\ref{e:y_equation_def}) therefore determine the mass $M$, the radius
$R$ and the quadrupole deformation $Y$ of a relativistic stellar
model.  The tidal deformability $\Lambda$ is then determined
algebraically from Eq.~(\ref{e:DefTidalDeformationAppendix}) with
$C=M/R$.  This third-order system of ordinary differential equations
to determine $M$ and $\Lambda$ is therefore more efficient to solve
numerically than the original fourth-order system,
Eqs.~(\ref{e:OVMass}), (\ref{e:OVPressure}), and
(\ref{e:ReggeWheeler}), derived by Hinderer~\cite{Hinderer2008,
  Hinderer2009}.

In our previous work on the inverse stellar structure
problem~\cite{Lindblom12}, we found that the Oppenheimer-Volkoff
equations could be solved more accurately and efficiently by
transforming them into a form that determines the mass $m(h)$ and
radius $r(h)$ as functions of the relativistic enthalpy $h$.  We use
this same transformation in this work to change
Eq.~(\ref{e:y_equation_def}) for the relativistic quadrupole
deformation $y(r)$ into an equation for $y(h)$.  The resulting
transformed stellar structure equations are,
\begin{eqnarray}
\frac{dm}{dh}&=&{\cal M}(m,r,\epsilon,p)\equiv
-\frac{4\pi r^3 \epsilon(r-2m)}{m+4\pi r^3 p},
\label{e:OVEquation_M}\\
\frac{dr}{dh}&=&{\cal R}(m,r,p)\equiv
-\frac{r(r-2m)}{m+4\pi r^3 p}
\label{e:OVEquation_R},\\
\frac{dy}{dh}&=&{\cal Y}(y,m,r,\epsilon,p,\Gamma)\equiv 
\frac{(r-2m)(y+1)y}{m+4\pi r^3p}+y\nonumber\\
&&+\frac{(m-4\pi r^3\epsilon)y}{m+4\pi r^3 p}
+\frac{4\pi r^3 (5\epsilon+9p)-6r}{m+4\pi r^3 p}\nonumber\\
&&+\frac{4\pi r^3(\epsilon+p)^2}{p\Gamma(m+4\pi r^3 p)}
-\frac{4(m+4\pi r^3 p)}{r-2m},
\label{e:dydhEquation}
\end{eqnarray}
where the quantities ${\cal M}(m,r,\epsilon,p)$, ${\cal R}(m,r,p)$ and
${\cal Y}(y,m,r,\epsilon,p,\Gamma)$ merely represent the expressions
on the right sides.  

The enthalpy based representation of the stellar structure
Eqs.~(\ref{e:OVEquation_M})--(\ref{e:dydhEquation}) are solved
numerically by specifying conditions, $m(h_c)=r(h_c)=0$ and
$y(h_c)=2$, at the center of the star where $h=h_c$ and then
integrating these equations out to the surface of the star where
$h=0$.  The right sides of
Eqs.~(\ref{e:OVEquation_M})--(\ref{e:dydhEquation}), i.e., the
functions ${\cal M}(m,r,\epsilon,p)$, ${\cal R}(m,r,p)$ and ${\cal
  Y}(y,m,r,\epsilon,p,\Gamma)$ are singular at the center of the star
$h=h_c$.  Consequently it is necessary to start any numerical
integration of these equations slightly away from that singular point.
The needed starting conditions can be obtained using a power series
solution to the equations.  The needed power series can be written in
the form,\footnote{We note that the power series expansion given in
  Eq.~(16) of Hinderer~\cite{Hinderer2008, Hinderer2009} for $H(r)$
  near $r=0$ contains a typographical error, which has been corrected
  in our derivation of Eqs.~(\ref{e:h_series_y}) and (\ref{e:y2Def}).}
\begin{eqnarray}
r(h) &=& r_1 (h_c - h)^{1/2} + r_3 (h_c -h )^{3/2} 
\nonumber\\
&&\quad+ {\cal O}(h_c - h)^{5/2},\label{e:h_series_r}\\
m(h) &=& m_3 (h_c - h)^{3/2} + m_5 (h_c - h)^{5/2} 
\nonumber\\
&&\quad+ {\cal O}(h_c - h)^{7/2},\label{e:h_series_m}\\
y(h) &=& 2 + y_2 (h_c - h) + {\cal O}(h_c-h)^2,
\label{e:h_series_y}
\end{eqnarray}
where $r_1$, $r_3$, $m_3$, $m_5$ and $y_2$ are given:
\begin{eqnarray}
r_1 &=& \left[\frac{3}{2\pi(\epsilon_c + 3 p_c)}\right]^{1/2},\\
r_3 &=& - \frac{r_1}{4(\epsilon_c+3p_c)}
\left[\epsilon_c - 3p_c -\frac{3(\epsilon_c+p_c)^2}{5 p_c \Gamma_c}\right],\\
m_3 &=& \frac{4\pi}{3} \epsilon_c r_1^3,\\
m_5 &=& 4\pi r_1^3\left[\frac{r_3\epsilon_c}{r_1} -
\frac{(\epsilon_c+p_c)^2}{5 p_c\Gamma_c}\right],\\
y_2&=&-\frac{6}{7(\epsilon_c+3p_c)}\left[
\frac{\epsilon_c}{3}+11 p_c +\frac{(\epsilon_c+p_c)^2}{p_c\Gamma_c}\right].
\qquad\label{e:y2Def}
\end{eqnarray}
The quantities $\epsilon_c$, $p_c$ and $\Gamma_c$ in these expressions
are the energy density, pressure and the adiabatic index evaluated at
the center of the star where $h=h_c$: $\epsilon_c=\epsilon(h_c)$,
$p_c=p(h_c)$, and $\Gamma_c =\Gamma(h_c)$.  We obtain the total mass
$M(h_c,\gamma_k)$ and dimensionless tidal deformation
$\Lambda(h_c,\gamma_k)$ by solving
Eqs.~(\ref{e:OVEquation_M})--(\ref{e:dydhEquation}) numerically
starting at $h=h_c$ using an equation of state with spectral
parameters $\gamma_k$.  The total mass is simply the surface value
$M(h_c,\gamma_k)=m(0)$ of this solution, while $\Lambda(h_c,\gamma_k)$
is determined from Eq.~(\ref{e:DefTidalDeformationAppendix}) using the
surface values $C=m(0)/r(0)$ and $Y=y(0)$.

It will be useful for our least-squares minimization problem to know
how the solutions to
Eqs.~(\ref{e:OVEquation_M})--(\ref{e:dydhEquation}) change as the
parameters $h_c$ and $\gamma_k$ are varied.  Let $\eta$ denote any
one of the parameters: $\eta=\{h_c,\gamma_k\}$.  We wish to derive
equations for the derivatives of the solutions to these equations with
respect to these parameters: $\partial m/\partial\eta$, $\partial
r/\partial\eta$ and $\partial h/\partial\eta$.  It is
straightforward to determine the needed auxiliary equations by
differentiating, Eqs.~(\ref{e:OVEquation_M})--(\ref{e:dydhEquation})
with respect to $\eta$:
\begin{eqnarray}
\frac{d}{dh}\left(\frac{\partial m}{\partial\eta}\right)&=&
\frac{\partial {\cal M}}{\partial m}\frac{\partial m}{\partial\eta}
+\frac{\partial {\cal M}}{\partial r}\frac{\partial r}{\partial\eta}
+\frac{\partial {\cal M}}{\partial \epsilon}\frac{\partial \epsilon}{\partial\eta}\nonumber \\
&&+\frac{\partial {\cal M}}{\partial p}\frac{\partial p}{\partial\eta},
\label{e:VarOVEquation_m}\qquad\\
\frac{d}{dh}\left(\frac{\partial r}{\partial\eta}\right)&=&
\frac{\partial {\cal R}}{\partial m}\frac{\partial m}{\partial\eta}
+\frac{\partial {\cal R}}{\partial r}\frac{\partial r}{\partial\eta}
+\frac{\partial {\cal R}}{\partial p}\frac{\partial p}{\partial\eta},
\qquad\qquad
\label{e:VarOVEquation_r}\\
\frac{d}{dh}\left(\frac{\partial y}{\partial\eta}\right)&=&
\frac{\partial {\cal Y}}{\partial y}\frac{\partial y}{\partial\eta}
+\frac{\partial {\cal Y}}{\partial m}\frac{\partial m}{\partial\eta}
+\frac{\partial {\cal Y}}{\partial r}\frac{\partial r}{\partial\eta}
\nonumber \\
&&
+\frac{\partial {\cal Y}}{\partial \epsilon}\frac{\partial \epsilon}{\partial\eta}
+\frac{\partial {\cal Y}}{\partial p}\frac{\partial p}{\partial\eta}
+\frac{\partial {\cal Y}}{\partial \Gamma}\frac{\partial \Gamma}{\partial\eta}.
\label{e:VarLoveNumber_y}
\end{eqnarray}
The various derivatives $\partial {\cal M}/\partial m$, etc. are
determined directly from the stellar structure equations,
Eqs.~(\ref{e:OVEquation_M})--(\ref{e:dydhEquation}):

\begin{eqnarray}
\frac{\partial {\cal M}}{\partial m}&=& \frac{8\pi r^3\epsilon 
- {\cal M}}{m+4\pi r^3 p}
,\\
\frac{\partial {\cal M}}{\partial r}&=& 
-4\pi r^2\frac{3p{\cal M}+2\epsilon(2r-3m)}{m+4\pi r^3 p}
,\\
\frac{\partial {\cal M}}{\partial p}&=& 
-\frac{4\pi r^3 {\cal M}}{m+4\pi r^3 p}
,\\
\frac{\partial {\cal M}}{\partial \epsilon}&=& 
-\frac{4\pi r^3(r-2m)}{m+4\pi r^3 p}
,
\end{eqnarray}

\begin{eqnarray}
\frac{\partial {\cal R}}{\partial m}&=& \frac{2r - {\cal R}}{m+4\pi r^3 p}
,\\
\frac{\partial {\cal R}}{\partial r}&=& 
-\frac{12\pi r^2p{\cal R}+2(r-m)}{m+4\pi r^3 p}
,\\
\frac{\partial {\cal R}}{\partial p}&=& 
-\frac{4\pi r^3 {\cal R}}{m+4\pi r^3 p},
\end{eqnarray}

\begin{eqnarray}
\frac{\partial {\cal Y}}{\partial y} &=& 1 
+ \frac{(r-2m)2y+r-m-4\pi r^3\epsilon}{m+4\pi r^3 p}
,\\
\frac{\partial {\cal Y}}{\partial m} &=& 
-\frac{(r-2m)(y+1)y}{(m+4\pi r^3p)^2}
-\frac{(2y+1)y}{m+4\pi r^3p}-\frac{4}{r-2m}\nonumber\\
&&-\frac{(m-4\pi r^3\epsilon)y}{(m+4\pi r^3 p)^2}
-\frac{4\pi r^3 (5\epsilon+9p)-6r}{(m+4\pi r^3 p)^2}\nonumber\\
&&-\frac{4\pi r^3(\epsilon+p)^2}{p\Gamma(m+4\pi r^3 p)^2}
-\frac{8(m+4\pi r^3 p)}{(r-2m)^2},
\end{eqnarray}

\begin{eqnarray}
\frac{\partial {\cal Y}}{\partial r} &=& 
\frac{(y+1)y}{m+4\pi r^3 p}
-\frac{12\pi r^2p(r-2m)(y+1)y}{(m+4\pi r^3p)^2}\nonumber\\
&&
-\frac{12\pi r^2 p(m-4\pi r^3\epsilon)y}{(m+4\pi r^3 p)^2}
-\frac{12\pi r^2\epsilon y}{m+4\pi r^3 p}\nonumber\\
&&
+\frac{12\pi r^2 (5\epsilon+9p)-6}{m+4\pi r^3 p}
+\frac{12\pi r^2(\epsilon+p)^2}{p\Gamma(m+4\pi r^3 p)}\nonumber\\
&&
-\frac{12\pi r^2p[4\pi r^3 (5\epsilon+9p)-6r]}{(m+4\pi r^3 p)^2}
-\frac{48\pi r^2 p}{r-2m}\nonumber\\
%\end{\eqnarray}
%\vfill\break
% 
%\begin{eqnarray}
&&-\frac{48\pi^2 r^5(\epsilon+p)^2}{\Gamma(m+4\pi r^3 p)^2}
+\frac{4(m+4\pi r^3 p)}{(r-2m)^2},
\end{eqnarray}

\begin{eqnarray}
\frac{\partial {\cal Y}}{\partial \epsilon} &=& 
\frac{4\pi r^3(5-y)}{m+4\pi r^3 p}
+\frac{8\pi r^3(\epsilon+p)}{p\Gamma(m+4\pi r^3 p)}
,\\ 
\frac{\partial {\cal Y}}{\partial p} &=& 
-\frac{4\pi r^3y[(r-2m)(y+1)
+m-4\pi r^3\epsilon]}{(m+4\pi r^3p)^2}\nonumber\\
&&
-\frac{4\pi r^3[4\pi r^3 (5\epsilon+9p)-6r]}{(m+4\pi r^3 p)^2}
+\frac{36\pi r^3}{m+4\pi r^3 p}\nonumber\\
&&
-\frac{16\pi^2r^6(\epsilon+p)^2}{p\Gamma(m+4\pi r^3 p)^2}
+\frac{8\pi r^3(\epsilon+p)}{p\Gamma(m+4\pi r^3 p)}\nonumber\\
&&
-\frac{4\pi r^3(\epsilon+p)^2}{p^2\Gamma(m+4\pi r^3 p)}
-\frac{16\pi r^3}{r-2m}
,\\ 
\frac{\partial {\cal Y}}{\partial \Gamma} &=& 
-\frac{4\pi r^3(\epsilon+p)^2}{p\Gamma^2(m+4\pi r^3 p)}.
\end{eqnarray}

For the case when $\eta=\gamma_k$, the derivatives
$\partial\epsilon/\partial\gamma_k$, $\partial p/\partial\gamma_k$ 
and $\partial \Gamma/\partial\gamma_k$
are determined from the equations that determine the spectral
representation of the equation of state.  The needed expressions are
given by,
\begin{eqnarray}
\frac{\partial\tilde\mu(h)}{\partial\gamma_k}&=&
\int_{h_0}^h\left[\log\left(\frac{h'}{h_0}\right)\right]^k
\frac{e^{h'}dh'}{\Gamma(h')},\\
\frac{\partial p(h)}{\partial\gamma_k}&=& -p(h)
\int_{h_0}^h\frac{\partial\tilde\mu(h')}{\partial\gamma_k}
\frac{e^{h'}dh'}{\left[\tilde\mu(h')\right]^2},
\label{e:dpdgammak}\\
\frac{\partial \epsilon(h)}{\partial\gamma_k}&=& 
\frac{\partial p(h)}{\partial\gamma_k}\frac{\epsilon(h)}{p(h)}-
\frac{\partial\tilde\mu(h)}{\partial\gamma_k}
\frac{e^{h}p(h)}{\left[\tilde\mu(h)\right]^2},
\label{e:depsilondgammak}\\
\frac{\partial \Gamma(h)}{\partial\gamma_k}&=& 
\left[\log\left(\frac{h}{h_0}\right)\right]^k\Gamma(h).
\label{e:dGammadgammak}
\end{eqnarray}
The integrals needed to determine these quantities can be performed
accurately and efficiently using Gaussian quadrature.  The equation of
state does not depend on the parameter $h_c$, and so
$\partial\epsilon/\partial h_c=\partial p/\partial
h_c=\partial\Gamma/\partial h_c=0$.  Consequently the equations that
determine $\partial m/\partial h_c$, $\partial r/\partial h_c$ and
$\partial y/\partial h_c$ in
Eqs.~(\ref{e:VarOVEquation_m})--(\ref{e:VarLoveNumber_y}) are somewhat
simpler than those for $\partial m/\partial\gamma_k$, $\partial
r/\partial\gamma_k$ and $\partial y/\partial\gamma_k$.

The functions $\partial m/\partial \eta$, $\partial r/\partial
\eta$ and $\partial y/\partial \eta$ are determined by solving
Eqs.~(\ref{e:VarOVEquation_m})--(\ref{e:VarLoveNumber_y}) numerically.
This can be done by integrating them from the center of the star where
$h=h_c$ out to the surface of the star where $h=0$.  To do this we
need to impose the appropriate boundary conditions for these functions
at $h=h_c$.  The needed boundary conditions can be found by
differentiating the power series solutions,
Eqs.~(\ref{e:h_series_r})--(\ref{e:h_series_y}) with respect to the
parameters $\eta$.  The quantities $r_1$, $r_3$, $m_3$, $m_5$, and
$y_2$ which appear in these power series solutions, depend on the
central values of the thermodynamic quantities
$\epsilon_c=\epsilon(h_c)$, $p_c= p(h_c)$, and $\Gamma_c=\Gamma(h_c)$,
and through them the parameters $\eta=\{h_c,\gamma_k\}$.  For the
case where $\eta=\gamma_k$ these derivatives can be written as
\begin{eqnarray}
&&\!\!\!\!\!
\frac{\partial r(h)}{\partial\gamma_k} =
\left[\frac{\partial r_1}{\partial \epsilon_c}
\frac{\partial\epsilon_c}{\partial\gamma_k}
+\frac{\partial r_1}{\partial p_c}
\frac{\partial p_c}{\partial\gamma_k}\right](h_c-h)^{1/2}\nonumber\\
&&\quad+\left[\frac{\partial r_3}{\partial \epsilon_c}
\frac{\partial\epsilon_c}{\partial\gamma_k}
+\frac{\partial r_3}{\partial p_c}
\frac{\partial p_c}{\partial\gamma_k}
+\frac{\partial r_3}{\partial \Gamma_c}
\frac{\partial \Gamma_c}{\partial\gamma_k}\right](h_c-h)^{3/2}\nonumber\\
&&\quad+ {\cal O}(h_c - h)^{5/2},
\label{e:h_series_dr_gammak}\\
&&\!\!\!\!\!
\frac{\partial m(h)}{\partial\gamma_k} =
\left[\frac{\partial m_3}{\partial \epsilon_c}
\frac{\partial\epsilon_c}{\partial\gamma_k}
+\frac{\partial m_3}{\partial p_c}
\frac{\partial p_c}{\partial\gamma_k}\right](h_c-h)^{3/2}\nonumber\\
&&\quad+\left[\frac{\partial m_5}{\partial \epsilon_c}
\frac{\partial\epsilon_c}{\partial\gamma_k}
+\frac{\partial m_5}{\partial p_c}
\frac{\partial p_c}{\partial\gamma_k}
+\frac{\partial m_5}{\partial \Gamma_c}
\frac{\partial \Gamma_c}{\partial\gamma_k}\right](h_c-h)^{5/2}\nonumber\\
&&\quad+ {\cal O}(h_c - h)^{7/2}.
\label{e:h_series_dm_gammak}\\
&&\!\!\!\!\!
\frac{\partial y(h)}{\partial\gamma_k} =
\left[\frac{\partial y_2}{\partial \epsilon_c}
\frac{\partial\epsilon_c}{\partial\gamma_k}
+\frac{\partial y_2}{\partial p_c}
\frac{\partial p_c}{\partial\gamma_k}
+\frac{\partial y_2}{\partial \Gamma_c}
\frac{\partial \Gamma_c}{\partial\gamma_k}
\right](h_c-h)\nonumber\\
&&\quad+ {\cal O}(h_c - h)^{2}.
\label{e:h_series_dy_gammak}
\end{eqnarray}
The derivatives of $r_1$, $r_3$, $m_3$, $m_5$ and $y_2$ with respect
to the parameters $\epsilon_c$, $p_c$ and $\Gamma_c$ which appear in
Eqs.~(\ref{e:h_series_dr_gammak})--(\ref{e:h_series_dy_gammak}) are
given by,
\begin{eqnarray}
\frac{\partial r_1}{\partial \epsilon_c}&=&
-\frac{r_1}{2(\epsilon_c+3p_c)},
\label{e:dr1_depsilonc}\\
\frac{\partial r_1}{\partial p_c}&=&
3\frac{\partial r_1}{\partial \epsilon_c}.\\
\frac{\partial r_3}{\partial \epsilon_c}&=&
\frac{r_3}{r_1}\frac{\partial r_1}{\partial \epsilon_c}
-\frac{r_1}{4(\epsilon_c+3p_c)}
\left[1 +\frac{4r_3}{r_1}
-\frac{6(\epsilon_c+3p_c)}{5p_c\Gamma_c}\right],\nonumber\\
&&\\
\frac{\partial r_3}{\partial p_c}&=&
\frac{r_3}{r_1}\frac{\partial r_1}{\partial p_c}
+\frac{3r_1}{4(\epsilon_c+3p_c)}
\left[1 -\frac{4r_3}{r_1}
-\frac{\epsilon_c^2-p_c^2}{5p_c^2\Gamma_c}\right],\nonumber\\
&&\\
\frac{\partial r_3}{\partial \Gamma_c}&=&
-\frac{3r_1(\epsilon_c+p_c)^2}{20p_c(\epsilon_c+3p_c)\Gamma_c^2},
\end{eqnarray}
\begin{eqnarray}
\frac{\partial m_3}{\partial \epsilon_c}&=&\frac{4\pi}{3}r_1^3\left[1+
\frac{3\epsilon_c}{r_1}\frac{\partial r_1}{\partial\epsilon_c}\right],\\
\frac{\partial m_3}{\partial p_c}&=&4\pi\epsilon_cr_1^2
\frac{\partial r_1}{\partial p_c},\\
\frac{\partial m_5}{\partial \epsilon_c}&=&4\pi r_1^2\left[r_3
+\frac{2\epsilon_cr_3}{r_1}\frac{\partial r_1}{\partial\epsilon_c}
+\epsilon_c\frac{\partial r_3}{\partial\epsilon_c}\right]\nonumber\\
&&-\frac{4\pi r_1^2(\epsilon_c+p_c)}{5p_c\Gamma_c}
\left[
2r_1+3(\epsilon_c+p_c)\frac{\partial r_1}{\partial\epsilon_c}\right],
\qquad\\
\frac{\partial m_5}{\partial p_c}&=&4\pi \epsilon_c r_1^2\left[
\frac{2r_3}{r_1}\frac{\partial r_1}{\partial p_c}+
\frac{\partial r_3}{\partial p_c}\right]\nonumber\\
&&+\frac{4\pi r_1^3(\epsilon_c+p_c)}{5p_c^2\Gamma_c}
\left[
\epsilon_c -\frac{3p_c(\epsilon_c+p_c)}{r_1}
\frac{\partial r_1}{\partial p_c}\right],\qquad\quad
\label{e:dm5_dpc}
\\
\frac{\partial m_5}{\partial \Gamma_c}&=&
4\pi r_1^3\left[\frac{\epsilon_c}{r_1}
\frac{\partial r_3}{\partial\Gamma_c}+
\frac{(\epsilon_c+p_c)^2}{5 p_c\Gamma_c^2}\right],
\label{e:dm5_dGammac}
\end{eqnarray}
\begin{eqnarray}
\frac{\partial y_2}{\partial \epsilon_c}&=&-\frac{y_2}{\epsilon_c+3p_c}
%\frac{6}{7(\epsilon_c+3p_c)^2}\left[
%\frac{\epsilon_c}{3}+11 p_c +\frac{(\epsilon_c+p_c)^2}{p_c\Gamma_c}\right]
%\nonumber\\
%&&
-\frac{6}{7(\epsilon_c+3p_c)}\left[
\frac{1}{3}+\frac{2(\epsilon_c+p_c)}{p_c\Gamma_c}\right],
\label{e:dy2_depsilonc}
\qquad\\
\frac{\partial y_2}{\partial p_c}&=&
-\frac{3y_2}{\epsilon_c+3p_c}
%\frac{18}{7(\epsilon_c+3p_c)^2}\left[
%\frac{\epsilon_c}{3}+11 p_c +\frac{(\epsilon_c+p_c)^2}{p_c\Gamma_c}\right]
%\nonumber\\
%&&
-\frac{6}{7(\epsilon_c+3p_c)}\left[
11 -\frac{\epsilon_c^2-p_c^2}{p_c^2\Gamma_c}\right],
\label{e:dy2_dpc}
\\
\frac{\partial y_2}{\partial \Gamma_c}&=&
\frac{6(\epsilon_c+p_c)^2}{7(\epsilon_c+3p_c)p_c\Gamma_c^2}.
\label{e:dy2_dGammac}
\end{eqnarray}
The values of the derivatives $\partial p_c/\partial\gamma_k$,
$\partial \epsilon_c/\partial\gamma_k$ and $\partial
\Gamma_c/\partial\gamma_k$ are obtained by evaluating
Eqs.~(\ref{e:dpdgammak})--(\ref{e:dGammadgammak}) at $h=h_c$.

For the case where $\eta=h_c$ the expressions for the derivatives
$\partial r/\partial \eta$, $\partial m/\partial \eta$ and
$\partial y/\partial \eta$ have somewhat different forms because
$h_c$ appears explicitly in the expansions in
Eqs.~(\ref{e:h_series_r})--(\ref{e:h_series_y}).  Differentiating
these series with respect to $h_c$, keeping only the leading terms,
gives
\begin{eqnarray}
&&\frac{\partial r(h)}{\partial h_c} =
\frac{r_1}{2}(h_c-h)^{-1/2}\nonumber\\
&&\qquad+
\left[\frac{\partial r_1}{\partial \epsilon_c}
\frac{\partial\epsilon_c}{\partial h_c}
+\frac{\partial r_1}{\partial p_c}
\frac{\partial p_c}{\partial h_c}+\frac{3 r_3}{2}
\right](h_c-h)^{1/2}\nonumber\\
&&\qquad+ {\cal O}(h_c - h)^{3/2},\label{e:h_series_dr_hc}\\
&&\frac{\partial m(h)}{\partial h_c} =
\frac{3 m_3}{2}(h_c-h)^{1/2}\nonumber\\
&&\qquad+\left[\frac{\partial m_3}{\partial \epsilon_c}
\frac{\partial\epsilon_c}{\partial h_c}
+\frac{\partial m_3}{\partial p_c}
\frac{\partial p_c}{\partial h_c}
+\frac{5 m_5}{2}\right](h_c-h)^{3/2}\nonumber\\
&&\qquad+ {\cal O}(h_c - h)^{5/2}.
\label{e:h_series_dm_hc}\\
&&\frac{\partial y(h)}{\partial h_c} = y_2 + {\cal O}(h_c-h).
\label{e:h_series_dy_hc}
\end{eqnarray}
The derivatives of $r_1$, $r_3$, $m_3$, and $m_5$ with respect to the
parameters $\epsilon_c$, $p_c$ which appear in
Eqs.~(\ref{e:h_series_dr_hc}) and (\ref{e:h_series_dm_hc}) are given
as before by the expressions in
Eqs.~(\ref{e:dr1_depsilonc})--(\ref{e:dm5_dpc}), while the derivatives
$\partial\epsilon_c/\partial h_c$ and $\partial p_c/\partial h_c$ are
obtained directly from the definitions of the enthalpy and the
adiabatic index at $h=h_c$:
\begin{eqnarray}
\frac{\partial p_c}{\partial h_c} &=& \epsilon_c + p_c,\label{e:dpcdhc}\\
\frac{\partial\epsilon_c}{\partial h_c} &=& \frac{(\epsilon_c + p_c)^2}
{p_c\Gamma(h_c)}.
\label{e:depsiloncdhc}
\end{eqnarray}

The discussion to this point has shown how to evaluate the derivatives
of $M$, $R$ and $Y$ with respect to the parameters
$\eta=\{h_c,\gamma_k\}$.  The quantity of primary interest in the
discussion of the inverse stellar structure problem in
Sec.~\ref{s:TidalDeformability} is the tidal deformability $\Lambda$.
Its derivatives are determined by those of $M$, $R$ and $Y$:
\begin{eqnarray}
\frac{\partial\Lambda}{\partial\eta} &=&
\frac{\partial\Lambda}{\partial C}
\left(\frac{C}{M}\frac{\partial M}{\partial \eta}
-\frac{C}{R}\frac{\partial R}{\partial \eta}\right)
+\frac{\partial\Lambda}{\partial Y}\frac{\partial Y}{\partial \eta}.
\qquad
\label{e:DLambdaDeta}
\end{eqnarray}
The derivatives $\partial\Lambda/\partial C$ and
$\partial\Lambda/\partial Y$ are given by,

\begin{eqnarray}
&&\!\!\!\!\!
\frac{\partial \Lambda}{\partial C} = \frac{2(Y-1)\Lambda}{2+2C(Y-1) -Y}
-\frac{4\Lambda}{1-2C}
-\frac{\Lambda}{\Xi}\frac{\partial\,\Xi}{\partial C},\qquad\\
&&\!\!\!\!\!
\frac{\partial \Lambda}{\partial Y}=\frac{(2C-1)\Lambda}{2+2C(Y-1)-Y}
-\frac{\Lambda}{\Xi}\frac{\partial\,\Xi}{\partial Y},
\end{eqnarray}
where
\begin{eqnarray}
\frac{\partial\,\Xi}{\partial C}&=&
4C^2[39-33Y+4C(3Y-2)+10C^2(1+Y)]\nonumber\\
&&-12(1-2C)[2-Y+2C(Y-1)]\log(1-2C)\nonumber\\
&&+6(1-2C)^2(Y-1)\log(1-2C)\nonumber\\
&&-6(1-2C)[2-Y+2C(Y-1)]\nonumber\\
&&+6[2-Y+2C(5Y-8)],\\
\frac{\partial\,\Xi}{\partial Y}&=&
4C^3(3C-11+2C^2)+2C(15C-3)\nonumber\\
&&-3(1-2C)^3\log(1-2C).
\end{eqnarray}

In summary, the macroscopic stellar properties $M$, $R$ and $Y$ are
determined by solving the stellar structure
Eqs.~(\ref{e:OVEquation_M})--(\ref{e:dydhEquation}). The dimensionless
tidal deformability $\Lambda$ is then determined algebraically from
them using Eq.~(\ref{e:DefTidalDeformationAppendix}).  The derivatives
of these properties $\partial M/\partial \eta$, $\partial R/\partial
\eta$, and $\partial Y/\partial \eta$ with respect to the parameters
$\eta=\{h_c,\gamma_k\}$ are determined by solving the perturbed
stellar structure
Eqs.~(\ref{e:VarOVEquation_m})--(\ref{e:VarLoveNumber_y}).  The
derivatives of the dimensionless tidal deformability
$\partial\Lambda/\partial\eta$ are then determined algebraically
from them using Eq.~(\ref{e:DLambdaDeta}).

%%%%%%%%%%%%%%%%%%%%%%%%%%%%%%%%%%%%%%%%%%%%%%%%%%%%%%%%%%%%%%%%%%%%%%%%%%%%%%%%
%%%%%%%%%%%%%%%%%%%%%%%%%%%%%%%%%%%%%%%%%%%%%%%%%%%%%%%%%%%%%%%%%%%%%%%%%%%%%%%%
\bibstyle{prd} 
\bibliography{References}
%%%%%%%%%%%%%%%%%%%%%%%%%%%%%%%%%%%%%%%%%%%%%%%%%%%%%%%%%%%%%%%%%%%%%%%%%%%%%%%%
\end{document}